\documentclass[12pt,journal,onecolumn]{IEEEtran}
\usepackage{graphicx}          
\usepackage{mathrsfs}

\usepackage{graphicx,float}
\usepackage{color}
\usepackage{amsmath,amsfonts,amssymb}
\usepackage{cite}
\usepackage{alltt}

\usepackage{makecell}
\usepackage{epstopdf}

\begin{document}

\newtheorem{theorem}{Theorem}
\newtheorem{lemma}{Lemma}
\newtheorem{proposition}{Proposition}
\newtheorem{definition}{Definition}
\newtheorem{remark}{Remark}
\newtheorem{corollary}{Corollary}


\title{\LARGE Distributed System Identification for Linear Stochastic Systems with Binary Sensors
\thanks{fukewei15@mails.ucas.ac.cn; hfchen@iss.ac.cn; wxzhao@amss.ac.cn}
\thanks{This work was supported by the National Key Research and Development Program of China under Grant 2018YFA0703800, the National Nature Science Foundation of China under Grant with No. 61822312 and the Strategic Priority Research Program of Chinese Academy of Sciences under Grant with No. XDA27000000.}
}


\author{\normalsize Kewei Fu$^{*,**}$,~~Han-Fu Chen$^{*,**}$,~~Wenxiao Zhao$^{*,**}$\\
\small
$^*$Key Laboratory of Systems and Control, Academy of Mathematics and Systems Science, Chinese
Academy of Sciences, Beijing 100190, China.\\
School of Mathematical Sciences, University of Chinese Academy of Sciences, Beijing 100049, China.
}

\date{}
\maketitle
\vspace{-1cm}

\begin{abstract}
The problem of distributed identification of linear stochastic system with unknown coefficient $\theta^*$ over time-varying networks is considered. For estimating $\theta^*$, each agent in the network can only access the input and the binary-valued output of the local system. Compared with the existing works on distributed optimization and estimation, the binary-valued local output observation considered in the paper makes the problem challenging. By assuming that the agent in the network can communicate with its adjacent neighbours, a stochastic approximation based distributed identification algorithm is proposed, and the consensus and convergence of the estimates are established. Finally, a numerical example is given showing that the simulation results are consistent with the theoretical analysis.

\end{abstract}

\IEEEpeerreviewmaketitle


\section{Introduction}

In recent years, the wireless sensor networks (WSN) \cite{akyildiz2007survey,Yick2008wireless} have received much attention from researchers of diverse areas, including consensus seeking \cite{jadbabaie2003coordination,ren2005consensus, olfati2007consensus,feng2019output}, multi-agent optimization \cite{nedic2009distributed,nedic2010constrained,simonetto2014distributed}, resource allocation (RA) \cite{yi2016initialization,nedic2018improved}, and multi-unmanned aerial vehicle (MUAV) control \cite{kuriki2014consensus} etc. For WSNs, usually,  there is no central node where the collected data can be processed, but there are a number of senors, called agents, which have limited capacity in computation, observation, and information communication. The agents in WSNs are required to cooperatively accomplish a global objective by using their local observations and information obtained from communication with their adjacent neighbors. Over the centralized approach, the distributed approach has the advantages in robustness on network link failure, in privacy protection, and in reduction on communication and computation cost, see, e.g., \cite{jadbabaie2003coordination,ren2005consensus,nedic2009distributed,nedic2010constrained,simonetto2014distributed,olfati2007consensus,feng2019output}.

In this paper, we consider the distributed identification of linear stochastic systems with unknown vector coefficient $\theta^*$ over time-varying networks with binary sensors. Each agent in the network is aimed at estimating $\theta^*$, but it can only access the input and the binary-valued output of the local system. Due to the limited sensing and computing capacity, each agent cannot identify $\theta^*$ by using its local observations only, while for identifiability it needs to exchange information with its adjacent neighbors.


The distributed identification problem has been studied in many works. The diffusion least mean square (LMS) algorithm is proposed in \cite{cattivelli2009diffusion,lopes2008diffusion}, where the constant step-sizes are adopted and the mean-square errors of estimates are derived by assuming that the regressors are mutually independent and Gaussian. The distributed LMS algorithm is also considered in \cite{schizas2009distributed}, where the bounds for the estimation errors are obtained. The stochastic approximation (SA) based distributed identification algorithms are introduced in \cite{stankovic2010decentralized,zhang2012distributed,lei2015distributed}, where the consistent estimates are derived for $\theta^*$ by assuming stationarity of the observed data and connectivity of the network. In all of the above works the traditional sensors are equipped in the network. By this we mean that the sensors generate continuously varying observations. However, for recent years, the quantized or binary-valued observations have attracted much attention in systems and control community, since such kind of sensors are usually with low complexity and with much less operational cost in comparison with the traditional ones, and hence, they are more attractive for applications \cite{guo2011adaptive}. In most of these works the centralized estimation and control problems with quantized or binary-valued observations are concerned, see \cite{diao2018event,zhao2017recursive,zhao2007identification,zhao2010identification} and references therein. The distributed consensus of multi-agent systems with quantized communications is studied in \cite{zhang2013quantized,WangZhangZhao}. However, to the best of our knowledge, the distributed identification problem with binary-valued sensors has not been discussed yet.

In this paper, the distributed identification of linear stochastic systems over time-varying networks with binary-valued sensors is considered. First, the local excitation conditions on each agent in the network are introduced to guarantee identifiability of the unknown coefficient $\theta^*$ of the system. Second,
by transforming the identification task to a root-searching problem, a distributed identification algorithm is introduced. Each agent relies only on the input, the binary-valued output of the local system, and the information derived from its adjacent neighbors. Third, it is proved that the estimates generated by the distributed algorithm are of both consensus and convergence with probability one. Finally, a numerical example is given demonstrating that the simulation results are consistent with the theoretical analysis.

The rest of the paper is organized as follows. The problem formulation and the distributed identification algorithm are given in Section 2. The assumptions are introduced in Section 3 and the main results are given in Section 4. A numerical example is presented in Section 5. Some concluding remarks are addressed in Section 6. The basic convergence result for DSAAWET and a convergence result for mixing random series to be used in the paper are given in Appendix.

\begin{table}[htbp]
\caption{Notations}
\centering
\begin{tabular}{l|p{5cm}<{\centering}}
     \hline
     $||v||,||A||$  &  $L_2$ norm of vector $v$, matrix $A$ \\
     \hline
     $\textbf{I}_m$  &  $m\times m$ identity matrix\\ \hline
     $\textbf{1}$  &  Vector or matrix with all entries equal to 1\\ \hline
     $\textbf{0}$  &  Vector or matrix with all entries equal to 0\\ \hline
     $X^T$  &  Transpose of matrix X\\ \hline
     $\mathrm{col}\{x_1,\dots,x_m\}$ & $\mathrm{col}\{x_1,\dots,x_m\}\triangleq[x_1^T,\dots,x_m^T]^T$\\
     \hline
     $\mathbb{I}_{A}(x)$  &  Indicator function, $\mathbb{I}_{A}(x)=1$ if $x\in A$, $\mathbb{I}_{A}(x)=0$ otherwise\\ \hline
     $\otimes$  &  Kronecker product\\
     \hline
     $\mathbb{E}[\cdot]$  &  Expectation operator\\ \hline
     $D_{\perp}$  &  $D_{\perp}\triangleq(\textbf{I}_N-\frac{\textbf{1}\textbf{1}^T}{N})\otimes\textbf{I}_l$ with $N$  the number of agents in the network and $l$ the dimension of coefficient vector\\
     \hline
     $\mathrm{sgn}(\cdot)$  &  $\mathrm{sgn}(x)=1$ if $x\geq0$, $\mathrm{sgn}(x)=0$ otherwise\\
     \hline
     $\mathbf{e}_i$  &  Vector with $i$-th entry being $1$ and others being zero\\
     \hline
\end{tabular}
\end{table}

\section{Problem Formulation and Distributed Identification Algorithm}

We first recall some basic concepts in graph theory which will be used in the paper. A time-independent digraph $\mathcal{G}=\{\mathcal{V},\mathcal{E}\}$ is called {\em strongly connected } if for any $i,j\in\mathcal{V}$,
there exists a directed path from \emph{i} to \emph{j}. A {\em directed path} is a sequence of edges $(i,i_1),(i_1,i_2),\dots,(i_{p-1},j)$
in the digraph with distinct agents $i_k\in\mathcal{V},~0\le k\le p-1$, where \emph{p} is called the length of this path.
A nonnegative matrix \emph{A} is called {\em doubly stochastic} if $A\textbf{1}=\textbf{1}$ and $\textbf{1}^TA=\textbf{1}^T$.

In the paper we consider a network with \emph{N} agents. The interaction relationship among agents is described by a time-varying digraph
$\mathcal{G}(k)=\{\mathcal{V},\mathcal{E}(k)\}$, where $k$ is the time, $\mathcal{V}=\{1,\dots ,N\}$ is the agent set, and $\mathcal{E}(k)\subset\mathcal{V}\times\mathcal{V}$ is the edge set. By
$(i,j)\in\mathcal{E}(k)$ we mean that agent $j$ can receive information from agent $i$ at time $k$. Assume
$(i,i)\in\mathcal{E}(k)$ for any $ k=1,2,\dots$ Denote the neighbors of agent
$i$ at time $k$ by $N_i(k)=\{j\in\mathcal{V}:(j,i)\in\mathcal{E}(k)\}$. The adjacent matrix associated
with the graph is denoted by $W(k)=[w_{ij}(k)]_{i,j=1}^N$, where $w_{ij}(k)>0$ if and only if
$(j,i)\in\mathcal{E}(k)$, otherwise $w_{ij}(k)=0$.

The dynamics of agent $i,~i=1,\cdots,N$ is given by
\begin{align}\label{1}
&y_{i,k+1}=\phi_{i,k}^T\theta^*+d_{i,k+1},
\end{align}
where $\theta^*\in\mathbb{R}^{l\times 1}$ is unknown for all agents, $\phi_{i,k}\in\mathbb{R}^{l\times 1}$, $y_{i,k+1}\in\mathbb{R}$, and $d_{i,k+1}\in\mathbb{R}$ are the input vector, output, and noise of agent $i$, respectively.
The measured output of agent $i$ is given by a binary sensor
\begin{align}
z_{i,k+1}=\mathbb{I}_{[y_{i,k+1}<c_{i,k}]},\label{2}
\end{align}
where $\{c_{i,k}\}_{k\ge 1}$ is a sequence of time-varying thresholds, which can online be tuned and will be specified later on.
For each agent $i\in \mathcal{V}$, the problem of distributed identification is to estimate $\theta^*$ by using its local input sequence $\{\phi_{i,k}\}$, the binary-valued measurements $\{z_{i,k+1}\}$, and the information obtained from exchange with its adjacent neighbors.

Note that the gradient of the $L_1$ minimization $\mathbb{E}|y_{i,k+1}-\phi_{i,k}^T\theta|$ is $\mathbb{E}[-\phi_{i,k}\mathrm{sgn}(y_{i,k+1}-\phi_{i,k}^T\theta)]=\mathbb{E}[-\phi_{i,k}(1-2\mathbb{I}_{[y_{i,k+1}<\phi_{i,k}^T\theta]})]$. For the above identification problem, an intuitive distributed algorithm would be:
\begin{align}
\theta_{i,k+1}\!=\!\sum_{j\in N_i(k)}w_{ij}(k)\theta_{j,k}
\!+\!\frac{1}{k}\phi_{i,k}(1\!-\! 2\mathbb{I}_{[y_{i,k+1}<\phi^T_{i,k}\theta_{i,k}]})\label{2'}
\end{align}
for all $i\in \mathcal{V}$, where the first term on the right hand is deemed as a consensus term while the second term is a gradient descent and the indicators $\{\mathbb{I}_{[y_{i,k+1}<\phi^T_{i,k}\theta_{i,k}]}\}_{k\geq1}$ are in fact the binary-valued output measurements of agent $i$ with the time-varying thresholds $\{c_{i,k}=\phi^T_{i,k}\theta_{i,k}\}_{k\geq1}$. Algorithm (\ref{2'}) is in fact a stochastic approximation algorithm (SAA). For the classical approaches for theoretical analysis of SAA, c.f., the ordinary differential equation method, some {\it a prioir} assumptions are required, such as the estimation sequence being bounded, see, e.g., \cite{Chen}. In order to avoid such assumptions, here we introduce a modified version of \eqref{2'} by using the expanding truncation technique in \cite{lei2019distributed,Chen}.

The distributed identification algorithm in this paper is given by:
\begin{align}
\sigma_{i,0}=&0,~\hat{\sigma}_{i,k}=\max_{j\in N_i(k)}\sigma_{j,k},\label{6}\\
\nonumber
\theta_{i,k+1}^{\prime}=&\Big{\{}\sum_{j\in N_i(k)}w_{ij}(k)\theta_{j,k} \mathbb{I}_{[\sigma_{j,k}=\hat{\sigma}_{i,k}]}\\
&\!+\!\frac{1}{k}\phi_{i,k}(1\!-\!2\mathbb{I}_{[y_{i,k+1}<\phi^T_{i,k}\theta_{i,k}]})\Big{\}} \cdot\mathbb{I}_{[\sigma_{i,k}=\hat{\sigma}_{i,k}]},\label{3}\\
\theta_{i,k+1}=&\theta_{i,k+1}^{\prime}\mathbb{I}_{[||\theta_{i,k+1}^{\prime}||\le\hat{\sigma}_{i,k}]},\label{4}\\
\sigma_{i,k+1}=&\hat{\sigma}_{i,k}+\mathbb{I}_{[||\theta_{i,k+1}^{\prime}||>\hat{\sigma}_{i,k}]},\label{5}
\end{align}
where $\{\theta_{i,k}\}_{k\geq1}$ is the estimation sequence for $\theta^*$ at agent $i$.


Noting that $
1-2\mathbb{I}_{[y_{i,k+1}<\phi^T_{i,k}\theta_{i,k}]}=\mathrm{sgn}(y_{i,k+1}-\phi^T_{i,k}\theta_{i,k})
$, the algorithm (\ref{3}) can be rewritten as follows:
\begin{align}
\nonumber
\theta_{i,k+1}^{\prime}&=\Big{\{}\sum_{j\in N_i(k)}w_{ij}(k)\theta_{j,k} \mathbb{I}_{[\sigma_{j,k}=\hat{\sigma}_{i,k}]}\\
&+\frac{1}{k}\phi_{i,k}\mathrm{sgn}(y_{i,k+1}-\phi^T_{i,k}\theta_{i,k})\Big{\}} \cdot\mathbb{I}_{[\sigma_{i,k}=\hat{\sigma}_{i,k}]}.\label{7}
\end{align}

\begin{remark} \label{Remark1}
Note that from the algorithm (\ref{6})--(\ref{5}), the estimate $\theta_{i,k+1}$ at agent $i$ only relies on its local observations $\{\phi_{i,k},z_{i,k}\}$ and the information $\theta_{j,k},j\in N_i(k)$ obtained from its neighbors. The algorithm (\ref{6})--(\ref{5}) is a DSAAWET given in \cite{lei2019distributed}, where for expanding truncations the sequence of positive numbers increasingly diverging to infinity is denoted by $\{M_k\}_{k\geq1}$, but here it is taken as $\{M_k=k\}_{k\geq1}$. So in (\ref{4}) and (\ref{5}) $M_{\widehat{\sigma}_{i,k}}$ turns to be $\widehat{\sigma}_{i,k}$.
By introducing the expanding truncation technique into the algorithm, the conditions for its convergence can significantly be relaxed, for example, the martingale difference sequence (MDS) conditions on observation noises \cite{Robbins} and the boundedness assumption on the estimation sequence \cite{Ljung}, being not required.
\end{remark}


\section{Assumptions on Network Systems}

The following assumptions are imposed on the network systems.

\begin{itemize}
\item[C1)] For each agent $i\in\mathcal{V}$, $\{\phi_{i,k}\}_{k\ge1}$ is strictly stationary with the density function $q_i(\cdot)$ and is an $\alpha$-mixing with mixing coefficients $\{\alpha(k)\}_{k\ge1}$ satisfying $\alpha(k)\le C\rho_1^k$ for some $C>0$ and $0<\rho_1<1$. Further, $\sup_k||\phi_{i,k}||\leq M<\infty$ for some constant $M>0$.

\item[C2)] The matrix $\mathbb{E}[\sum_{i=1}^N\phi_{i,k}\phi_{i,k}^T]$ is positive definite for all $k$.

\item[C3)] For any $i\in\mathcal{V}$, $\{d_{i,k}\}$ is a sequence of independent and identically distributed (iid) random variables with the distribution function $F_{i,d}(\cdot)$ and the density function $f_{i,d}(\cdot)$. {\color{black} Further, it is assumed that $F_{i,d}(0)=\frac{1}{2}$, $f_{i,d}(\cdot)$ is continuous, $f_{i,d}(0)>0$}, and the sequences $\{\phi_{i,k}\}$ and $\{d_{i,k+1}\}$ are mutually independent.

\item[C4)] For the time-varying network $(\mathcal{V},\mathcal{E}(k))$, the following conditions are assumed.\\
a) The adjacent matrices $W(k)$ are doubly stochastic for each $k\geq 0$;\\
b) There exists a constant $0<\kappa<1$ such that
$
w_{ij}(k)\ge\kappa,
$
whenever $j\in N_i(k)$ for all $i\in\mathcal{V}$ and $k\geq 0$;\\
c) The digraph $\mathcal{G}_{\infty}=\{\mathcal{V},\mathcal{E}_{\infty}\}$ is strongly connected with
$\mathcal{E}_{\infty}=\{(j,i):(j,i)\in\mathcal{E}(k)$ for infinitely many indices of $k\};$\\
d) There exists a positive integer $B$ such that
\begin{gather}\nonumber
(j,i)\in\mathcal{E}(k)\cup\mathcal{E}(k+1)\cup\cdots\cup\mathcal{E}(k+B-1)
\end{gather}
for any $(j,i)\in\mathcal{E}_{\infty}$ and any $k\ge 1$.
\end{itemize}

\begin{remark}\label{Remark3}
If C4) holds, then by Proposition 1 given in \cite{nedic2009distributed}, there exist constants $c>0$ and $0<\rho_2<1$ such that
\begin{gather}
\left\|\Phi(k,s)-\frac{1}{N}\textbf{1}\textbf{1}^T\right\|\le c\rho_2^{k-s+1}\quad\forall~k\ge s,\label{8}
\end{gather}
where
\begin{gather}\nonumber
\Phi(k,s)\triangleq W(k)\cdots W(s)~ \forall~k\ge s~~ \mathrm{and}~~\Phi(k,k+1)\triangleq\textbf{I}_N.
\end{gather}
\end{remark}


Set
\begin{align}
&y_{k}\triangleq\mathrm{col}\{y_{1,k},\dots,y_{N,k}\}\in\mathbb{R}^{N\times 1},\label{9}\\
&\phi_{k}\triangleq[\phi_{1,k},\dots,\phi_{N,k}]\in\mathbb{R}^{l\times N},\label{10}
\end{align}
\begin{align}
&f(\theta)\triangleq\sum_{i=1}^N\mathbb{E} [\phi_{i,k}\text{sgn}(y_{i,k+1}-\phi_{i,k}^T\theta)], \label{11}\\
&f_i(\theta)\triangleq\mathbb{E}[\phi_{i,k} \text{sgn}(y_{i,k+1}-\phi_{i,k}^T\theta)],\label{12}\\
&O_{i,k+1}\triangleq\phi_{i,k} \mathrm{sgn}(y_{i,k+1}-\phi^T_{i,k}\theta_{i,k}),\label{34}
\end{align}
and
\begin{align}
\nonumber\epsilon_{i,k+1}\triangleq&\phi_{i,k}\text{sgn}(y_{i,k+1}-\phi_{i,k}^T\theta_{i,k}) \\
&-\mathbb{E}[\phi_{i,k}\text{sgn}(y_{i,k+1}-\phi_{i,k}^T \theta_{i,k})]. \label{35}
\end{align}
By C1) and C3), the functions $f(\theta)$ and $f_i(\theta)$ are free of time $k$.


Before proving strong consistency of the algorithm (\ref{6})--(\ref{5}), we show the following technical result.


\begin{lemma}\label{Lemma1}
Assume that C1)-C3) hold. Then $\theta^*$ is the unique zero of $f(\theta)$:
$
f(\theta^*)=0.
$
\end{lemma}

\emph{\textbf{Proof}}: We first show $f(\theta^*)=0$. By (\ref{11}) and (\ref{1}), we have
\begin{align}\nonumber
&f(\theta)=\sum_{i=1}^N\mathbb{E}[\phi_{i,k}\text{sgn}(y_{i,k+1}-\phi_{i,k}^T\theta)]\\ \nonumber
&=\sum_{i=1}^N\mathbb{E}[\phi_{i,k}\text{sgn}(d_{i,k+1}-\phi_{i,k}^T(\theta-\theta^*))]\\
&=\sum_{i=1}^N\mathbb{E}[\phi_{i,k}\mathbb{E}[\text{sgn}(d_{i,k+1}-\phi_{i,k}^T(\theta-\theta^*))|\phi_{i,k}]].\label{13}
\end{align}
Since $\{\phi_{i,k}\}$ and $\{d_{i,k+1}\}$ are mutually independent by C3), it holds that
\begin{align}\nonumber
&\mathbb{E}[\text{sgn}(d_{i,k+1}-\phi_{i,k}^T(\theta-\theta^*))|\phi_{i,k}]\\ \nonumber
=&(\mathbb{E}[\text{sgn}(d_{i,k+1}-y^T(\theta-\theta^*))])\vert_{y=\phi_{i,k}}\\ \nonumber
=&\int_{\phi_{i,k}^T(\theta-\theta^*)}^{+\infty}f_{i,d}(x)dx-\int_{-\infty}^{\phi_{i,k}^T(\theta-\theta^*)}f_{i,d}(x)dx\\
=&1-2F_{i,d}(\phi_{i,k}^T(\theta-\theta^*)),\label{14}
\end{align}
where $F_{i,d}(\cdot)$ and $f_{i,d}(\cdot)$ are the distribution function and the density function of $d_{i,k}$, respectively.
Combining (\ref{13}) with (\ref{14}) we obtain
\begin{align}
f(\theta)=\sum_{i=1}^N\mathbb{E}[\phi_{i,k}(1-2F_{i,d}(\phi_{i,k}^T(\theta-\theta^*)))].\label{15}
\end{align}
From (\ref{15}) and C3) we know that $f(\theta^*)=0$. Next, we show that $\theta^*$ is the unique zero of $f(\cdot)$.

Define
\begin{gather}
G(\theta)\triangleq\mathbb{E}\|y_{k+1}-\phi_k^T\theta\|_1,\label{16}
\end{gather}
where $y_k$ and $\phi_k$ are given by (\ref{9}) and (\ref{10}).

As mentioned above $-f(\theta)$ is the gradient of $G(\theta)$ denoted by $\triangledown G(\theta)$ \cite{ChenYin}. Since $G(\theta)$ is convex and $\theta^*$ is a root of $\triangledown G(\theta)$, to prove the uniqueness of the root it suffices to show that the Hessian matrix of $G(\theta)$ is positive definite at $\theta^*$, or equivalently, to prove that the Jacobian matrix of $-f(\theta)$ is positive definite at $\theta^*$.

{\color{black} Calculating the Jacobian matrix of $-f(\theta)$ at $\theta^*$, by (\ref{15}) we have that
\begin{align}\nonumber
-\frac{\partial f(\theta)}{\partial \theta}\Big|_{\theta=\theta^*}&=\sum_{i=1}^N\mathbb{E}[\phi_{i,k}\phi_{i,k}^T\cdot2f_{i,d}(\phi_{i,k}^T(\theta-\theta^*))]\Big|_{\theta=\theta^*}\\ \nonumber
&=\mathbb{E}[\sum_{i=1}^N\phi_{i,k}\phi_{i,k}^T\cdot2f_{i,d}(\phi_{i,k}^T(\theta-\theta^*))]\Big|_{\theta=\theta^*}\\ \label{17}
&=\mathbb{E}[\sum_{i=1}^N\phi_{i,k}\phi_{i,k}^T\cdot2f_{i,d}(0)].
\end{align}
Set $a\triangleq\min_{i=1,\dots,N}f_{i,d}(0)$. By C3, $a>0$. From (\ref{17}) and by C2) it follows
\begin{gather}\nonumber
-\frac{\partial f(\theta)}{\partial \theta}\Big|_{\theta=\theta^*}\ge 2a\cdot \mathbb{E}[\sum_{i=1}^N\phi_{i,k}\phi_{i,k}^T]>0.
\end{gather}
Hence, $\theta^*$ is the unique zero of $f(\theta)$.\hfill$\square$


\begin{remark}\label{Remark5}
Assumption C2) requires that $\mathbb{E}[\sum_{i=1}^N\phi_{i,k}\phi_{i,k}^T]$ is positive definite, which is in fact an identifiability condition for $\theta^*$ in the distributed identification framework. It is clear that this condition does not ensure that $\theta^*$ is identifiable for each agent $i\in \mathcal{V}$ based on its local observations $\{\phi_{i,k},z_{i,k}\}$ only. To see this, let us consider a network system with $N=l$, i.e., the number of agents is equal to the dimension of unknown parameter $\theta^*$. For each $i\in \mathcal{V}$, let
$$
\begin{cases}
y_{i,k+1}=\phi_{i,k}^T\theta^{*}+d_{i,k+1}, \\
\phi_{i,k}=w_{i,k}\mathbf{e}_i,\\
z_{i,k+1}=\mathbb{I}_{[y_{i,k+1}<c_{i,k}]},
\end{cases}
$$
where $\{w_{i,k}\}_{k\geq1}$ is a sequence of iid random variables with zero mean and finite variance, and $\{w_{i,k}\}_{k\geq1}$ and $\{w_{j,k}\}_{k\geq1}$ are mutually independent for any $i\neq j$. It is directly verified that the matrix $\mathbb{E}[\sum_{i=1}^N\phi_{i,k}\phi_{i,k}^T]$ is positive definite, while for each agent $i$ based on $\{\phi_{i,k},z_{i,k+1}\}$, only the $i$-th elements of $\theta^*$ is identifiable.
\end{remark}

\section{Strong Consistency of Distributed Estimation Algorithm}

We have the following main result.

\begin{theorem}\label{Theorem1}
Assume C1)--C4) hold. Then for any $i\in\mathcal{V}$, the estimates $\{\theta_{i,k}\}_{k\geq1}$ generated by (\ref{6})--(\ref{5}) are of both consensus and convergence, i.e.,
\begin{gather}
\lim_{k\to\infty}\theta_{i,k}=\theta^* \quad \mathrm{a.s.}\quad\forall~i\in \mathcal{V}.\label{97}
\end{gather}
\end{theorem}

As indicated in Remark \ref{Remark1}, the algorithm (\ref{6})--(\ref{5}) is a DSAAWET \cite{lei2019distributed}.
However, the proof of Theorem \ref{Theorem1} is not a straightforward application of the general convergence result of DSAAWET. See \cite{lei2019distributed} or Theorem \ref{TheoremA1} in Appendix. We first establish Lemmas \ref{Lemma2}--\ref{Lemma7}.

\begin{itemize}

\item In Lemma \ref{Lemma2} we first analyze the truncation numbers of agents and show that the differences of truncation numbers among agents are bounded.

\item In Lemmas \ref{Lemma3}--\ref{Lemma4} we first show that there exists convergent subsequences $\{\theta_{i,n_k}\}_{k\geq1},i\in \mathcal{V}$ of $\{\theta_{i,k}\}_{k\geq1},i\in \mathcal{V}$ generated from the algorithms (\ref{6})--(\ref{5}), and then analyze the asymptotical properties of estimates among the convergent subsequences.

\item In Lemma \ref{Lemma5} we prove the noise condition required by assumption A4) of Theorem \ref{TheoremA1} given in Appendix.

\item In Lemma \ref{Lemma6} we show that the number of truncations in the network is finite and in Lemma \ref{Lemma7} we establish the consensus of the estimates.

\end{itemize}

Then based on Lemmas \ref{Lemma2}--\ref{Lemma7} we can prove Theorem \ref{Theorem1}.

Let us denote by $\tau_{i,m}\triangleq\inf\{k:\sigma_{i,k}=m\}$ the smallest time when the truncation number of agent $i$ reaches $m$, by $\tau_m\triangleq\min_{i\in\mathcal{V}}\tau_{i,m}$ the smallest time when at least one of agents whose truncation number has reached $m$, and by $\sigma_{k}\triangleq\max_{i\in\mathcal{V}}\sigma_{i,k}$ the largest truncation number among all agents at time $k$. Define $\tilde{\tau}_{j,m}\triangleq\min\{\tau_{j,m},\tau_{m+1}\}$.

The following lemma follows directly from \cite{lei2019distributed}.

\begin{lemma} \label{Lemma2}~~
\begin{itemize}
\item[i)] (Remark 3.1 in \cite{lei2019distributed}) For $\{\theta_{i,k}\}_{k\geq1},i\in \mathcal{V}$ generated by (3)-(6) with any initial values, it holds that:
\begin{gather}
\theta_{i,k+1}=\textbf{0}~~\text{when}~~\sigma_{i,k+1}>\sigma_{i,k}.\label{18}
\end{gather}
\item[ii)] (Lemma 4.2 in \cite{lei2019distributed}) Assume C4) holds. Then
\begin{gather}\nonumber
\tilde{\tau}_{j,m}\le\tau_m+B(N-1)~~\forall j\in\mathcal{V}~\mathrm{and}~m\ge0.
\end{gather}
\item[iii)] (Lemma 5.5 in \cite{lei2019distributed}) Assume C4) holds. If $\lim_{k\to\infty}\sigma_k=\sigma<\infty$, then there exists an integer $k_0>0$ such that
\begin{gather}\nonumber
\sigma_{j,k}=\sigma~~\forall j\in\mathcal{V}~\mathrm{and}~ k\ge k_0.
\end{gather}
\end{itemize}
\end{lemma}

Define
\begin{align*}
&\Theta_k\triangleq \mathrm{col}\{\theta_{1,k},\dots,\theta_{N,k}\},
\end{align*}
\begin{align*}
&\theta_k\triangleq\frac{1}{N}\sum_{i=1}^N\theta_{i,k},
\end{align*}
and
\begin{align*}
&\Theta_{\perp,k}\triangleq\Theta_k-(\textbf{1}\otimes I_l)\theta_k.
\end{align*}

\begin{lemma} \label{Lemma3}
If C1) and C4) hold, then for $\{\Theta_k\}_{k\geq1}$ generated by (\ref{6})--(\ref{5}), there exists some bounded subsequence $\{\Theta_{n_k}\}_{k\geq1}$ with $\sigma_{i,n_k}=\sigma_{n_k}$ for any $i\in\mathcal{V}$ and all sufficiently large $k$.
\end{lemma}

\emph{\textbf{Proof}}: The following analysis is carried out on a fixed sample path.

If $\lim_{k\to\infty}\sigma_k=\sigma<\infty$, then by iii) in Lemma \ref{Lemma2} there exists a positive integer $k_0$ such that there is no truncation for all agents in the network for $k\geq k_0$, i.e., $\sigma_{i,k}=\sigma$ for $k\ge k_0$ and $\forall i\in\mathcal{V}$, and hence the estimation sequence $\{\Theta_{k}\}_{k\geq1}$ is bounded.

Next, we consider the case where $\lim_{k\to\infty}\sigma_{i_0,k}=\infty$ for some $i_0\in \mathcal{V}$.

By C4) d) it follows that $i_0\in N_i(k)\bigcup N_i(k+1)N_i(k)\cdots \bigcup N_i(k+B-1)$ $\forall i\in \mathcal{V}$, say, $i_0\in N_i(k+l),~0\leq l\leq B-1$. Then $\widehat{\sigma}_{i,k+l}=\max\limits_{j\in N_i(k+l)}\sigma_{j,k+l}\geq \sigma_{i_0,k+l}$, and hence $\lim_{k\to\infty}\sigma_{i,k}=\infty~\forall i\in \mathcal{V}$ and $\lim_{k\to\infty}\sigma_k=\infty$.

Set $D\triangleq(N-1)B$ and $c_b\triangleq\sqrt{N}D$. It suffices to show that for all sufficiently large $m>m_0\triangleq D$,
\begin{align}
&\sigma_{i,\tau_m+D}=m~~\forall i\in\mathcal{V}\label{19}
\end{align}
and
\begin{align}
&||\Theta_{\tau_m+D}||\le c_b.\label{20}
\end{align}

Set $\{k\triangleq\tau_m\}_{m\geq1}$, which is, in fact, a subsequence of the positive numbers. For sufficiently large $m\ge m_0$ and any $q=1,\dots,D$, let us show that the following assertions take place:

\begin{itemize}
\item[1)] For any agent $i$ with $\sigma_{i,k}=m$ it holds that
\begin{gather}
\sigma_{i,k+q}=m~~\text{and}~~||\theta_{i,k+q}||\le q\le m.\label{21}
\end{gather}

\item[2)] For any agent $j$ with $\sigma_{j,k}<m$ it holds that
\begin{gather}
\sigma_{j,k+q}\le m\label{22}
\end{gather}
and
\begin{gather}
||\theta_{j,k+q}||\le q~~\text{whenever}~~\sigma_{j,k+q}=m.\label{23}
\end{gather}
\end{itemize}

We prove 1) and 2) by induction. We first show 1) for $q=1$.

Noting $k=\tau_m$, by the definition of $\tau_m$ we know that $\sigma_{j,k}\le m$ and $\sigma_{j,k-1}<m$ for any $j\in\mathcal{V}$. Noticing $\sigma_{i,k}=m$ for agent $i$ and (\ref{6}), we know that $\hat{\sigma}_{i,k}=m$. Since $\sigma_{i,k-1}<\sigma_{i,k}$, by (\ref{18}) it follows that $\theta_{i,k}=\textbf{0}$. Then from (\ref{3}) we obtain
\begin{align}
\nonumber\theta_{i,k+1}^{\prime}=&\Big{\{}\sum_{j\in N_i(k)}w_{ij}(k)\theta_{j,k} \mathbb{I}_{[\sigma_{j,k}=\hat{\sigma}_{i,k}]}\\
\nonumber&+\frac{1}{k}\phi_{i,k}(1-2\mathbb{I}_{[y_{i,k+1}<\phi^T_{i,k}\theta_{i,k}]})\Big{\}} \cdot\mathbb{I}_{[\sigma_{i,k}=\hat{\sigma}_{i,k}]}\\
\nonumber=&{}\sum_{j\in N_i(k)}w_{ij}(k)\theta_{j,k} \mathbb{I}_{[\sigma_{j,k}=\hat{\sigma}_{i,k}]}\\
\nonumber&+\frac{1}{k}\phi_{i,k}(1-2\mathbb{I}_{[y_{i,k+1}<\phi^T_{i,k}\theta_{i,k}]})\\
\nonumber=&{}\sum_{j\in N_i(k)}w_{ij}(k)\theta_{j,k} \mathbb{I}_{[\sigma_{j,k}=\hat{\sigma}_{i,k}]}\\
&+\frac{1}{k}\phi_{i,k}\text{sgn}(y_{i,k+1}-\phi^T_{i,k}\theta_{i,k}).\label{24}
\end{align}

For $j\in N_i(k)$, if $\sigma_{j,k}=\widehat{\sigma}_{i,k}=m$, then by noting that $\sigma_{j,k-1}<m$ and (\ref{18}), we have $\theta_{j,k}=\textbf{0}$, while if $\sigma_{j,k}<\widehat{\sigma}_{i,k}$, we have $\mathbb{I}_{[\sigma_{j,k}=\hat{\sigma}_{i,k}]}=0$. Hence from (\ref{24}) we obtain
\begin{align}
\theta_{i,k+1}^{\prime}=\frac{1}{k}\phi_{i,k}\text{sgn}(y_{i,k+1}-\phi^T_{i,k}\theta_{i,k}).\label{25}
\end{align}

Assumption C1) indicates that $\{\sup_k||\phi_{i,k}||\}_{k\geq1}$ is bounded and hence $\lim_{k\to\infty}\frac{1}{k}||\phi_{i,k}||=0$. So there exists a sufficiently large $k_0$ such that
\begin{gather}
\frac{1}{k}||\phi_{i,k}||<1~~\forall i\in\mathcal{V}~~\forall k\geq k_0.\label{26}
\end{gather}

Noticing $k=\tau_m\ge m$, $M_m=m$, and $m_0=D$, from (\ref{25})--(\ref{26}) it follows that for sufficiently large $m\ge (m_0\vee k_0)$
\begin{gather}
||\theta_{i,k+1}^{\prime}||\le\frac{1}{k}||\phi_{i,k}||\le 1\le m,\label{27}
\end{gather}
for agent $i$ with $\sigma_{i,k}=m$. Hence by (\ref{4}) and (\ref{5}) it follows that $\theta_{i,k+1}=\theta_{i,k+1}^{\prime}$ and $\sigma_{i,k+1}=\hat{\sigma}_{i,k}=m$, which by noticing (\ref{27}) guarantees that $||\theta_{i,k+1}||\le 1$. Thus we have proved 1) for $q=1$.

Next, we prove 2) for $q=1$.

For agent $j$ with $\sigma_{j,k}<m$, by definition of $k=\tau_m$ we know that $\hat{\sigma}_{j,k}\le m$.

If $\hat{\sigma}_{j,k}=m$, by noting $\sigma_{j,k}<m$, from (\ref{3}) it follows that $\theta_{j,k+1}^{\prime}=\textbf{0}$ and then from (\ref{5}) $\sigma_{j,k+1}=\hat{\sigma}_{j,k}=m$. If $\hat{\sigma}_{j,k}<m$, then from (\ref{5}) it follows that $\sigma_{j,k+1}\le\hat{\sigma}_{j,k}+1\le m$.
Thus in the case $\sigma_{j,k}<m$ we conclude that $\sigma_{j,k+1}\le m$.

Noting $\sigma_{j,k}<m$ and (\ref{18}), if $\sigma_{j,k+1}=m$, then we have that $\theta_{j,k+1}=\textbf{0}$. Therefore, 2) holds for $q=1$.

Next, we assume that 1) and 2) hold for $q=1,\dots,p$ with $p<D$. We now prove that they are also true for $q=p+1$.
We first consider case 1). From the inductive assumption we have $\sigma_{i,k+p}\leq m,~i\in \mathcal{V}$, and hence $\sigma_{k+p}\le m$. For the agent $i$ with $\sigma_{i,k+p}=m$, we have $\hat{\sigma}_{i,k+p}=m$. Then by (\ref{3}), we have
\begin{align}\nonumber
&\theta_{i,k+p+1}^{\prime}=\sum_{j\in N_{i}(k+p)}w_{ij}(k+p)\theta_{j,k+p}\mathbb{I}_{[\sigma_{j,k+p}=m]}\\
&\quad+\frac{1}{k+p}\phi_{i,k+p}\text{sgn}(y_{i,k+p+1}-\phi_{i,k+p}^T\theta_{i,k+p}).\label{28}
\end{align}

By the inductive assumption and (\ref{28}) and noticing that $W(k+p)$ is doubly stochastic, we have
\begin{gather}
||\theta_{i,k+p+1}^{\prime}||\le p+\frac{1}{k+p}||\phi_{i,k+p}||.\label{29}
\end{gather}

Similar to (\ref{26}) we know that for sufficiently large $k_0$,
\begin{gather}\nonumber
\frac{1}{k+p}||\phi_{i,k+p}||\le 1\quad\forall i\in\mathcal{V}~~\forall k\geq k_0,
\end{gather}
which incorporating with (\ref{29}) implies that for sufficiently large $m\ge (m_0\vee k_0)$
\begin{gather}
||\theta_{i,k+p+1}^{\prime}||\le p+1\le m=\widehat{\sigma}_{i,k+p}.\label{30}
\end{gather}
From the algorithm (\ref{4})--(\ref{5}) by (\ref{30}) we know that $\theta_{i,k+p+1}=\theta^{\prime}_{k+p+1}$ and $\sigma_{i,k+p+1}=\hat{\sigma}_{i,k+p}=m$. So, we have proved that 1) holds for $q=p+1$.

We now show that 2) holds for $q=p+1$.
By the inductive assumption we have $\sigma_{j,k+p}\le m$. For the case $\sigma_{j,k+p}=m$, similar to (\ref{28})--(\ref{30}), we can show that 2) holds for $q=p+1$ if $\sigma_{j,k+p}=m$.

For the case $\sigma_{j,k+p}<m$, we consider two cases: $\hat{\sigma}_{j,k+p}=m$ and $\hat{\sigma}_{j,k+p}<m$. For the case $\hat{\sigma}_{j,k+p}=m$, since $\sigma_{j,k+p}<\hat{\sigma}_{j,k+p}$, by (\ref{3}) we have that $\theta_{j,k+p+1}^{\prime}=\textbf{0}$ and by (\ref{5}) we have $\sigma_{j,k+p+1}=\hat{\sigma}_{j,k+p}=m$. For the case $\hat{\sigma}_{j,k+p}<m$, by (\ref{5}) we have that $\sigma_{j,k+p+1}\le\hat{\sigma}_{j,k+p}+1\le m$. Therefore, for the case $\sigma_{j,k+p}<m$ we have that $\sigma_{j,k+p+1}\le m$.

If $\sigma_{j,k+p+1}=m$ and $\sigma_{j,k+p}=m$, then by 1) for $q=p+1$, we know that $\|\theta_{i,k+q+1}\|\leq p+1$. If $\sigma_{j,k+p+1}=m$ and $\sigma_{j,k+p}<m$, then by (\ref{18}) we conclude $\theta_{j,k+p+1}=\textbf{0}$. Thus 2) holds for $q=p+1$.

We now have proved (\ref{21})--(\ref{23}), from which we conclude that $\tau_{i,m+1}>k+D$ for any $ i\in\mathcal{V}$ and all sufficiently large $m\ge (m_0\vee k_0)$. Since $k=\tau_m$, by the definition of $\tau_{m+1}$ we have $\tau_{m+1}-\tau_m>D$. By ii) in Lemma \ref{Lemma2} we obtain $\widetilde{\tau}_{i,m}=\min(\tau_{i,m},\tau_{m+1})\le\tau_m+D$ for any $ i\in\mathcal{V}$, and hence either $\tau_{i,m}\le\tau_m+D$, or $\tau_{m+1}\le\tau_m+D$. However, the last inequality is impossible, because we have proved that $\tau_{m+1}-\tau_m>D$. Hence we obtain $\tau_{i,m}\le\tau_m+D$ for any $i\in \mathcal{V}$, from which it follows that $\sigma_{i,\tau_m+D}\ge m$ by noticing $\sigma_{i,\tau_{i,m}}=m$. On the other hand, from $\tau_{i,m+1}>\tau_{m}+D$ it follows that $\sigma_{i,\tau_{m}+D}\le m$ for any $ i\in\mathcal{V}$. The above analysis yields that $\sigma_{i,\tau_m+D}=m$ for any $ i\in\mathcal{V}$.

Define the subsequence $\{\theta_{i,\tau_m+D}\}_{m\geq(m_0\vee k_0)}$ of $\{\theta_{i,k}\}_{k\geq1}$, $i\in \mathcal{V}$. From (\ref{21}) and (\ref{23}) it is seen that $||\theta_{i,\tau_m+D}||\le D~\forall i\in\mathcal{V}$ for all large enough $m\ge (m_0\vee k_0)$ and
\begin{gather}\nonumber
||\Theta_{\tau_m+D}||\le\sqrt{N}\max_{i}||\theta_{i,\tau_m+D}||\le\sqrt{N}D.
\end{gather}
This is (\ref{20}), and Lemma \ref{Lemma3} is proved.\hfill$\square$

The following result characterizes properties of $\{\Theta_m,m=n_k,\cdots,m(n_k,T)\}$ along the bounded subsequence $\{\Theta_{n_k}\}_{k\geq1}$ generated by the algorithm.

\begin{lemma}\label{Lemma4} Let $\{\Theta_{n_k}\}_{k\geq1}$ be a bounded subsequence generated by (\ref{6})--(\ref{5}) with $\sigma_{i,n_k}=\sigma_{n_k},~\forall i \in\mathcal{V}$. Assume that C1), C3), and C4) hold. Then there exist constants $c_1>0$, $c_2>0$, and $M_0'>0$ such that for sufficiently large $k$ and small enough $T>0$ ,
\begin{gather}
||\Theta_{m+1}-\Theta_{n_k}||\le c_1T+M_0^{\prime},\label{31}\\
||\theta_{m+1}-\theta_{n_k}||\le c_2T\label{32}
\end{gather}
for $m=n_k,\cdots, m(n_k,T)$ where $m(k,T)=\max\{m:\sum_{i=k}^m\frac{1}{k}\le T\}$.
\end{lemma}

\emph{\textbf{Proof}}: For simplicity of notations, we set
$$a_k\triangleq\frac{1}{k}\quad\mathrm{and}\quad O_{i,k+1}\triangleq\phi_{i,k}\text{sgn}(y_{i,k+1}-\phi_{i,k}^T\theta_{i,k}).$$

Since $\{\Theta_{n_k}\}_{k\geq1}$ is a bounded subsequence with $\sigma_{i,n_k}=\sigma_{n_k}~\forall i \in\mathcal{V}$, from (\ref{6})--(\ref{5}) we know $\sigma_{i,n_k}=\hat{\sigma}_{i,n_k}~\forall i\in\mathcal{V}$ and derive
\begin{gather}\nonumber
\theta^{\prime}_{i,n_k+1}=\sum_{j\in N_{i}(n_k)}w_{ij}(n_k)\theta_{j,n_k}+a_{n_k}O_{i,n_k+1}.
\end{gather}

If there is no truncation at time $n_k+1$ for any agent $i\in\mathcal{V}$, then
\begin{gather}\nonumber
\theta_{i,n_k+1}=\theta_{i,n_k+1}^{\prime}=\sum_{j\in N_{i}(n_k)}w_{ij}(n_k)\theta_{j,n_k}+a_{n_k}O_{i,n_k+1},
\end{gather}
and (\ref{6})--(\ref{5}) can be rewritten in the compact form:
\begin{align}\nonumber
\Theta_{n_k+s+1}=&(W(n_k+s)\otimes\textbf{I}_l)\Theta_{n_k+s}\\
&+a_{n_k+s}(F(\Theta_{n_k+s})+\epsilon_{n_k+s+1})\label{33}
\end{align}
with $s=0$, where $F(\Theta_{k})=\mathrm{col}\{f_{1}(\theta_{1,k}),\dots,f_N(\theta_{N,k})\}$, $\epsilon_{k}=\mathrm{col}\{\epsilon_{1,k},\dots,\epsilon_{N,k}\}$ with $f_i(\cdot)$ and $\epsilon_{i,k}$ defined by (\ref{12}) and (\ref{35}), respectively.

Since $\{\Theta_{n_k}\}_{k\geq1}$ is bounded, there exists a constant $C>0$ such that
\begin{gather}
||\Theta_{n_k}||\le C,~~k\ge 1.\label{36}
\end{gather}

Define
\begin{align}
&c_0\triangleq 2\sqrt{N}\max_{i\in\mathcal{V}}\mathbb{E}||\phi_{i,1}||+1,\label{37}\\
&M_0^{\prime}\triangleq 1+C(c\rho_2+2),\label{38}\\
&H_1\triangleq \max_{\Theta}\{||F(\Theta)||:||\Theta||\le M_0^{\prime}+1+C\},\label{39}\\
&c_1\triangleq H_1+c_0\left(3+\frac{c(\rho_2+1)}{1-\rho_2}\right)~~\mathrm{and}~~c_2\triangleq\frac{H_1+c_0}{\sqrt{N}},\label{40}
\end{align}
where the constants $c>0$ and $0<\rho_2<1$ are given in (\ref{8}).

Select $T>0$ small enough such that
\begin{gather}
c_1T<1.\label{41}
\end{gather}

For any $k\ge 1$, define
\begin{align}
\nonumber s_k\triangleq&\sup\{s:n_k \leq s\leq 2m(n_k,T)~\big|~|\Theta_j\!-\!\Theta_{n_k}||\le c_1T\!\!+\!\!M_0^{\prime},\\
&~n_k\leq j\leq s\}.\label{42}
\end{align}
It is clear that $s_k\geq n_k$, and from (\ref{36}) and (\ref{41}), for any $k\ge 1$ and $ n_k\le s\le s_{k}$
\begin{gather}
||\Theta_{s}||\le c_1T+M_0^{\prime}+||\Theta_{n_k}||\le M_0^{\prime}+1+C.\label{43}
\end{gather}

In the following we will show that $s_k>m(n_k,T)~\forall k\geq1$. Assume the converse: there exists a subsequence of $\{n_k,s_k\}$, for simplicity of notations, denoted still by $\{n_k,s_k\}$, such that
\begin{gather}
s_k\le m(n_k,T).\label{44}
\end{gather}

We first show that there exists an integer $k_1>1$ such that for all $k\ge k_1$
\begin{gather}
s_k<\tau_{\sigma_{n_k}+1}.\label{45}
\end{gather}

To prove (\ref{45}), we consider two cases:
1) $\lim_{k\to\infty}\sigma_k=\infty$ and 2) $\lim_{k\to\infty}\sigma_k=\sigma<\infty$.

For Case 1), since the truncation bounds $\{M_k\}$ used in DSAAWET is a sequence of positive numbers increasingly diverging to infinity as mentioned in Remark \ref{Remark1}, there exists a positive integer $k_1$ such that $M_{\sigma_{n_k}}>M^{\prime}_0+1+C$ for all $k\ge k_1$. Hence, from (\ref{43}) we know $s_k<\tau_{\sigma_{n_k}+1}$. For Case 2), since $\lim_{k\to\infty}\sigma_k=\sigma<\infty$, there exists a positive $k_1$ such that $\sigma_{n_k}=\sigma$ for all $k\ge k_1$, and hence $\tau_{\sigma_{n_k}+1}=\tau_{\sigma+1}=\infty$. This implies (\ref{45}).

From (\ref{45}) it follows that (\ref{33}) holds for $s:0\le s\le s_k-n_k-1$.

Next, we investigate the property of the noise sequence $\{\epsilon_{i,k+1}\}$. For $n_k\le s\le s_k$, we have
\begin{align}\nonumber
&\frac{1}{T}\Big\|\sum_{m=n_k}^sa_m\epsilon_{i,m+1}\Big\|\\ \nonumber
\le&\frac{1}{T}\Big\|\sum_{m=n_k}^s\frac{1}{m}[\phi_{i,k}\text{sgn}(y_{i,k+1}-\phi_{i,k}^T\theta_{i,k})\\ \nonumber
&-\mathbb{E}[\phi_{i,k}\text{sgn}(y_{i,k+1}-\phi_{i,k}^T\theta_{i,k})]]\Big\|\\ \nonumber
\le&\frac{1}{T}\sum_{m=n_k}^s\frac{1}{m}\|\phi_{i,m}\|+\frac{1}{T}\sum_{m=n_k}^s\frac{1}{m}\mathbb{E}\|\phi_{i,k}\|\\ \nonumber
\le&\frac{1}{T}\sum_{m=n_k}^{m(n_k,T)}\frac{1}{m}\|\phi_{i,m}\| +\frac{1}{T}\sum_{m=n_k}^{m(n_k,T)}\frac{1}{m}\mathbb{E}\|\phi_{i,k}\|\\ \nonumber
\le&\frac{1}{T}\sum_{m=n_k}^{m(n_k,T)}\frac{1}{m}[\|\phi_{i,m}\|-\mathbb{E}\|\phi_{i,m}\|]\\
&+2\cdot\frac{1}{T}\sum_{m=n_k}^{m(n_k,T)}\frac{1}{m}\mathbb{E}\|\phi_{i,k}\|.\label{46}
\end{align}

We first analyse $\sum_{m=n_k}^{m(n_k,T)}\frac{1}{m}[\|\phi_{i,m}\|-\mathbb{E}\|\phi_{i,m}\|]$. By C1), $\{\phi_{i,k}\}_{k\ge1}$ is an $\alpha$-mixing process. For any constant $\epsilon>0$, by boundedness of $\|\phi_{i,k}\|$ we have
\begin{align}\nonumber
&\sum_{m=1}^{\infty}\frac{1}{m^2}(\mathbb{E}|(\|\phi_{i,m}\| -\mathbb{E}\|\phi_{i,m}\|)|^{2+\epsilon})^{\frac{2}{2+\epsilon}}\\
=&O\left(\sum_{m=1}^{\infty}\frac{1}{m^2}\right)<\infty.\label{47}
\end{align}

From (\ref{47}) and by Theorem \ref{TheoremA2} in Appendix we have
\begin{gather}
\sum_{m=1}^{\infty}\frac{1}{m}[\|\phi_{i,m}\|-\mathbb{E}\|\phi_{i,m}\|]<\infty~~\mathrm{a.s.},\label{48}
\end{gather}
and hence
\begin{gather}
\lim_{k\to\infty}\sum_{m=n_k}^{m(n_k,T)}\frac{1}{m}[\|\phi_{i,m}\|-\mathbb{E}\|\phi_{i,m}\|]=0.\label{49}
\end{gather}

For the second term in (\ref{46}), we have
\begin{gather}
2\cdot\frac{1}{T}\sum_{m=n_k}^{m(n_k,T)}\frac{1}{m}\mathbb{E}\|\phi_{i,k}\|\le 2\mathbb{E}\|\phi_{i,k}\|.\label{50}
\end{gather}

Combining (\ref{46}), (\ref{49}), and (\ref{50}) we obtain
\begin{gather}
\limsup_{k\to\infty}\frac{1}{T}\|\sum_{m=n_k}^s\frac{1}{m}\epsilon_{i,m+1}\|\le2\mathbb{E}||\phi_{i,k}||,\label{51}
\end{gather}
and for sufficiently large $k\ge k_1$,
\begin{gather}
\frac{1}{T}\|\sum_{m=n_k}^s\frac{1}{m}\epsilon_{i,m+1}\|\le2\mathbb{E}\|\phi_{i,k}\|+\frac{1}{\sqrt{N}},\label{52}
\end{gather}
from which we conclude that for sufficiently large $k\ge k_1$,
\begin{gather}
\|\sum_{m=n_k}^sa_m\epsilon_{m+1}\|\le c_0T~~\forall s:n_k\le s \le s_k.\label{53}
\end{gather}

Define
\begin{gather}
Z_{s_k+1}\triangleq(W(s_k)\otimes\textbf{I}_l)\Theta_{s_k}+a_{s_k}(F(\Theta_{s_k})+\epsilon_{s_k+1})\label{54}
\end{gather}
and
\begin{align}
z_{s_k+1}\triangleq\frac{\textbf{1}^T\otimes\textbf{I}_l}{N}Z_{s_k+1}.\label{55}
\end{align}

It is clear that $Z_{s_k+1}$ coincides with $\Theta_{s_k+1}$ if there is no truncation at $s_k+1$.

Multiplying (\ref{33}) by $\frac{\mathbf{1}^T\otimes \mathbf{I}_l}{N}$ from left, by noticing $\frac{\textbf{1}^T\otimes\textbf{I}_l}{N}(W(s)\otimes\textbf{I}_l)=\frac{\textbf{1}^T\otimes\textbf{I}_l}{N}~\forall s\ge 0$ for any doubly stochastic $W(s)$, we derive
\begin{align*}
\theta_{n_k+s+1}=&\theta_{n_k+s}+a_{n_k+s}\frac{\mathbf{1}^T\otimes \mathbf{I}_l}{N}(F(\Theta_{n_k+s})+\epsilon_{n_k+s+1})
\end{align*}
for $0\leq s\leq s_k-n_k-1,$ and hence
\begin{align*}
\theta_{s_k}=&\theta_{n_k}+\frac{\mathbf{1}^T\otimes \mathbf{I}_l}{N}\sum\limits_{m=n_k}^{s_k-1}a_{m}(F(\Theta_{m})+\epsilon_{m+1}).
\end{align*}

Then, from (\ref{54}) and (\ref{55}) it follows that
\begin{align}\nonumber
z_{s_k+1}&=\theta_{s_k}+\frac{\textbf{1}^T\otimes\textbf{I}_l}{N}a_{s_k}(F(\Theta_{s_k})+\epsilon_{s_k+1})\\
&=\theta_{n_k}+\frac{\textbf{1}^T\otimes\textbf{I}_l}{N}\sum_{m=n_k}^{s_k}a_{m}(F(\Theta_{m})+\epsilon_{m+1}).\label{56}
\end{align}
From this, by (\ref{53}) and the definition of $s_k$, it follows that
\begin{align}\nonumber
&\|z_{s_k+1}-\theta_{n_k}\|\\ \nonumber
\le&\|\frac{\textbf{1}^T\otimes\textbf{I}_l}{N}\|\cdot\|\sum_{m=n_k}^{s_k}a_{m}(F(\Theta_{m})+\epsilon_{m+1})\|\\ \nonumber
\le&\frac{1}{\sqrt{N}}\sum_{m=n_k}^{s_k}a_m\|F(\Theta_m)\|+\frac{1}{\sqrt{N}}\|\sum_{m=n_k}^{s_k}a_m\epsilon_{m+1}\|\\
=&O(T)\label{57}
\end{align}
for all sufficiently large $k\ge k_1$.

By noticing $(W(k)\otimes \mathbf{I}_l)\cdots (W(s)\otimes \mathbf{I}_l)=\Phi(k,s)\otimes \mathbf{I}_l$ for $k\geq s$, from (\ref{33}) it follows that
\begin{align}
\nonumber&\Theta_{s_k}=(\Phi(s_k,n_k)\otimes \mathbf{I}_l)\Theta_{n_k}\\
&~~+\sum_{m=n_k}^{s_k}a_{m}(\Phi(s_k,m+1)\otimes \mathbf{I}_l)(F(\Theta_{m})+\epsilon_{m+1}).\label{57'}
\end{align}

Setting $Z_{\perp,s_k+1}\triangleq Z_{s_k+1}-(\textbf{1}\otimes\textbf{I}_l)z_{s_k+1}$, by (\ref{54}), (\ref{56}), and (\ref{57'}) we have
\begin{align}\nonumber
Z_{\perp,s_k+1}=&(W(s_k)\otimes \mathbf{I}_l)\Theta_{s_k}+a_{s_k}(F(\Theta_{s_k})+\epsilon_{s_k+1})\\
\nonumber&-\left(\frac{\mathbf{1} \mathbf{1}^T}{N}\otimes \mathbf{I}_l\right)(\Theta_{s_k}+a_{s_k}(F(\Theta_{s_k})+\epsilon_{s_k+1}))\\
\nonumber=&[(\Phi(s_k,n_k)-\frac{1}{N}\textbf{1}\textbf{1}^T)\otimes\textbf{I}_l]\Theta_{n_k}\\ \nonumber
&+\sum_{m=n_k}^{s_k}a_m[(\Phi(s_k,m+1)\!-\!\frac{1}{N}\textbf{1}\textbf{1}^T)\otimes\textbf{I}_l]F(\Theta_m)\\
&+\sum_{m=n_k}^{s_k}a_m[(\Phi(s_k,m+1)\!-\!\frac{1}{N}\textbf{1}\textbf{1}^T)\otimes\textbf{I}_l]\epsilon_{m+1},\label{58}
\end{align}
from which and by (\ref{8}), (\ref{36}) and (\ref{39}) it follows that
\begin{align}\nonumber
&\|Z_{\perp,s_k+1}\|\le Cc\rho_2^{s_k+1-n_k}+\sum_{m=n_k}^{s_k}a_mH_1c\rho_2^{s_k-m}\\
&~~~~+\|\sum_{m=n_k}^{s_k}a_m[(\Phi(s_k,m+1)-\frac{1}{N}\textbf{1}\textbf{1}^T)\otimes\textbf{I}_l]\epsilon_{m+1}\|.\label{59}
\end{align}

Let us estimate the last term on the right hand side of (\ref{59}). Set $\Gamma_n\triangleq\sum_{m=1}^na_m\epsilon_{m+1}$. By (\ref{53}) we derive
\begin{gather}
||\Gamma_s-\Gamma_{n_k-1}||\le c_0T ~~ \forall s:n_k\le s \le s_k.\label{60}
\end{gather}
We have the following equalities,
\begin{align}\nonumber
&\sum_{m=n_k}^sa_m(\Phi(s,m+1)\otimes\textbf{I}_l)\epsilon_{m+1}\\ \nonumber
=&\sum_{m=n_k}^s(\Phi(s,m+1)\otimes\textbf{I}_l)(\Gamma_m-\Gamma_{m-1})\\ \nonumber
=&\sum_{m=n_k}^s(\Phi(s,m+1)\otimes\textbf{I}_l)(\Gamma_m-\Gamma_{n_k-1})\\ \nonumber
&\quad-\sum_{m=n_k}^s(\Phi(s,m+1)\otimes\textbf{I}_l)(\Gamma_{m-1}-\Gamma_{n_k-1}),
\end{align}
from which it follows that
\begin{align}\nonumber
&\|\sum_{m=n_k}^sa_m(\Phi(s,m+1)\otimes\textbf{I}_l)\epsilon_{m+1}\|\\ \nonumber
\le&\|\Gamma_s-\Gamma_{n_k-1}\|\\
\nonumber&+\sum_{m=n_k}^{s-1}\|\Phi(s,m+1)-\Phi(s,m+2)\|\cdot\|\Gamma_m-\Gamma_{n_k-1}\|\\ \nonumber
\le& c_0T+\sum_{m=n_k}^{s-1}(c\rho_2^{s-m}+c\rho_2^{s-m-1})c_0T\\
\le& c_0T+\frac{c(\rho_2+1)}{1-\rho_2}c_0T\quad \forall s:n_k\le s\le s_k.\label{61}
\end{align}
From (\ref{53}) and (\ref{61}) it follows that
\begin{align}\nonumber
&\|\sum_{m=n_k}^sa_m[(\Phi(s,m+1)-\frac{1}{N}\textbf{1}\textbf{1}^T)\otimes\textbf{I}_l]\epsilon_{m+1}\|\\
\le&(2+\frac{c(\rho_2+1)}{1-\rho_2})c_0T    \quad \text{for sufficiently large}~k\ge k_1.\label{62}
\end{align}
From (\ref{59}) and (\ref{62}) we further obtain
\begin{gather}
\|Z_{\perp,s_k+1}\|\le Cc\rho_2+1+(2+\frac{c(\rho_2+1)}{1-\rho_2})c_0T\label{63}
\end{gather}
for sufficiently large $k\ge k_1$.

Since $Z_{s_k+1}=Z_{\perp,s_k+1}+(\textbf{1}\otimes\textbf{I}_l)z_{s_k+1}$, we derive
\begin{align}\nonumber
&\|Z_{s_k+1}-\Theta_{n_k}\|\\ \nonumber
=&\|(\textbf{1}\otimes\textbf{1}_l)z_{s_k+1}+Z_{\perp,s_k+1} -\Theta_{\perp,n_k}-(\textbf{1}\otimes\textbf{I}_l)\theta_{n_k}\|\\
\le&\|Z_{\perp,s_k+1}\|+\|\Theta_{\perp,n_k}\|+\sqrt{N}\|z_{s_k+1}-\theta_{n_k}\|.\label{64}
\end{align}

Noticing $||\Theta_{\perp,n_k}||\le 2C~\forall k\ge 1$, from (\ref{57}) and (\ref{63}) we know that for sufficiently large $k\ge k_1$
\begin{align}\nonumber
&\|Z_{s_k+1}-\Theta_{n_k}\|\\ \nonumber
\le& Cc\rho_2+1+(2+\frac{c(\rho_2+1)}{1-\rho_2})c_0T+\sqrt{N}\frac{H_1+c_0}{\sqrt{N}}T+2C\\
=&c_1T+M_0^{\prime},\label{65}
\end{align}
where $M_0^{\prime}$ and $c_1$ are defined by (\ref{37}) and (\ref{38}). Therefore,
\begin{gather}\nonumber
||Z_{s_k+1}||\le||\Theta_{n_k}||+M_0^{\prime}+c_1T\le M_0^{\prime}+1+C.
\end{gather}

This means that for the algorithm (\ref{6})--(\ref{5}), there is no truncation at $s_k+1$ for sufficiently large $k\ge k_1$. Therefore, (\ref{33}) holds for $s=s_k-n_k$ and $\Theta_{s_k+1}=Z_{s_k+1}$ for sufficiently large $k\ge k_1$. Hence, from (\ref{65}) we obtain
\begin{gather}\nonumber
||\Theta_{s_k+1}-\Theta_{n_k}||\le M_0^{\prime}+c_1T,
\end{gather}
which contradicts with the definition of $s_k$ in (\ref{42}). Thus we have proved that $s_k>m(n_k,T)$ for sufficiently large $k\ge k_1$. Consequently, from (\ref{42}) we know that (\ref{31}) holds for sufficiently large $k$.

Since $s_k>m(n_k,T)$, similar to (\ref{57}) it can be proven that (\ref{32}) holds for sufficiently large $k$.
This finishes the proof.\hfill$\square$

For the observation noises $\{\epsilon_{i,k}\}$ defined by (\ref{35}), the following result takes place.


\begin{lemma}\label{Lemma5}
If C1), C3), and C4) hold, then for any convergent subsequence $\{\Theta_{n_k}\}_{k\geq1}$ of $\{\Theta_{k}\}_{k\geq1}$ with $\sigma_{i,n_k}=\sigma_{n_k}~\forall i\in\mathcal{V}$, it holds that
\begin{gather}
\lim_{j\to\infty}\limsup_{k\to\infty}\frac{1}{T_j} \Big\|\sum_{m=n_k}^{m(n_k,T_j)}\frac{1}{m}\epsilon_{m+1}\Big \|=0~~\forall T_j\in [0,T].\label{66}
\end{gather}
\end{lemma}

\emph{\textbf{Proof}}: Since $\{\Theta_{n_k}\}$ is a convergent subsequence, by definition we know that $\{\theta_{n_k}\}$ is also convergent. Denote by $\bar{\theta}$ the limit of $\{\theta_{n_k}\}$, i.e., $\theta_{n_k}\to\bar{\theta}$ as $k\to\infty$. Let $\{T_j\}_{j\ge1}$ be a sequence of positive numbers tending to zero with $T_j>T_{j+1}$. Let $\{\lambda_n\}$ be a nonincreasing sequence of positive numbers with $\lambda_n\to 0$ as $n\to\infty$ such that $\|\theta_{n_k}-\bar{\theta}\|<\frac{\lambda_{n_k}}{2}$. Denote by $\mathcal{S}$ a countable set dense in $\mathbb{R}^l$. Let $\{\theta(n)\}_{n\ge1}\subset\mathcal{S}$ be a sequence satisfying $\|\theta(n)-\bar{\theta}\|<\frac{\lambda_n}{2}$.

We rewrite the noise $\epsilon_{i,m+1}$ as follow:
\begin{gather}
\epsilon_{i,m+1}\!=\!\epsilon_{i,m+1}^{(1)}(n)\!+\!\epsilon_{i,m+1}^{(2)}(n)\!+\!\epsilon_{i,m+1}^{(3)}(n) \!+\!\epsilon_{i,m+1}^{(4)},\label{67}
\end{gather}
where
\begin{align}\nonumber
&\epsilon_{i,m+1}^{(1)}(n)=\phi_{i,m}\text{sgn}(y_{i,m+1}-\phi_{i,m}^T\theta_{i,m})\\
&\qquad\qquad\quad -\phi_{i,m}\text{sgn}(y_{i,m+1}-\phi_{i,m}^T\theta(n)),\label{68}\\\nonumber
&\epsilon_{i,m+1}^{(2)}(n)=\phi_{i,m}\text{sgn}(y_{i,m+1}-\phi_{i,m}^T\theta)\Big|_{\theta=\theta(n)}\\
&\qquad\qquad\quad-\mathbb{E}[\phi_{i,m}\text{sgn}(y_{i,m+1}-\phi_{i,m}^T\theta)]\Big|_{\theta=\theta(n)},\label{69}\\ \nonumber
&\epsilon_{i,m+1}^{(3)}(n)=\mathbb{E}[\phi_{i,m}\text{sgn}(y_{i,m+1}-\phi_{i,m}^T\theta)]\Big|_{\theta=\theta(n)}\\
&\qquad\qquad\quad-\mathbb{E}[\phi_{i,m}\text{sgn}(y_{i,m+1}-\phi_{i,m}^T\theta)]\Big|_{\theta=\overline{\theta}},\label{70}
\end{align}
and
\begin{align}
\nonumber
&\epsilon_{i,m+1}^{(4)}=\mathbb{E}[\phi_{i,m}\text{sgn}(y_{i,m+1}-\phi_{i,m}^T\theta)]\Big|_{\theta=\overline{\theta}}\\
&\qquad\qquad-\mathbb{E}[\phi_{i,m}\text{sgn}(y_{i,m+1}-\phi_{i,m}^T\theta)]\Big|_{\theta=\theta_{i,m}}.\label{71}
\end{align}

To prove the lemma it suffices to verify (\ref{66}) with $\epsilon_{m+1}$ replaced by $\epsilon_{i,m+1}^{(h)}$, $h=1,\cdots,4$. We first consider the case $h=1$. From the definition of $\epsilon_{i,m+1}^{(1)}(n)$, it follows that
\begin{align}
\nonumber
&\frac{1}{T_j}\Big\|\sum_{m=n_k}^{m(n_k,T_j)}\frac{1}{m}\phi_{i,m}\Big[\text{sgn}(y_{i,m+1}-\phi_{i,m}^T\theta_{i,m})\\
\nonumber
&-\text{sgn}(y_{i,m+1}-\phi_{i,m}^T\theta(n))\Big]\Big\|\\
\nonumber
=&\frac{1}{T_j}\Big\|\sum_{m=n_k}^{m(n_k,T_j)}\frac{1}{m}\phi_{i,m}\Big[1-2\mathbb{I}_{[y_{i,m+1}<\phi_{i,m}^T\theta_{i,m}]}\\ \nonumber
&\quad-1+2\mathbb{I}_{[y_{i,m+1}<\phi_{i,m}^T\theta(n)]}\Big]\Big\|\\ \nonumber
=&\frac{2}{T_j}\Big\|\sum_{m=n_k}^{m(n_k,T_j)}\frac{1}{m}\phi_{i,m}\Big[\mathbb{I}_{[y_{i,m+1}<\phi_{i,m}^T\theta_{i,m}]}\\
&\quad-\mathbb{I}_{[y_{i,m+1}<\phi_{i,m}^T\theta(n)]}\Big]\Big\|.\label{72}
\end{align}

For $m=n_k,\cdots,m(n_k,T_j)$, a direct calculation leads to
\begin{align}\nonumber
&|\mathbb{I}_{[y_{i,m+1}<\phi_{i,m}^T\theta_{i,m}]}-\mathbb{I}_{[y_{i,m+1}<\phi_{i,m}^T\theta(n)]}|
\end{align}
\begin{align}
\nonumber
\le&\mathbb{I}_{[y_{i,m+1}<\phi_{i,m}^T\theta_{i,m},y_{i,m+1}\ge\phi_{i,m}^T\theta(n)]}\\ \nonumber
&\quad+\mathbb{I}_{[y_{i,m+1}\ge\phi_{i,m}^T\theta_{i,m}, y_{i,m+1}<\phi_{i,m}^T\theta(n)]}\\
\nonumber
=&\mathbb{I}_{\big[\phi_{i,m}^T(\theta_{i,m}-\theta(n))>y_{i,m+1}-\phi_{i,m}^T\theta(n)\ge0\big]}\\ \nonumber
&\quad+\mathbb{I}_{\big[0>y_{i,m+1}-\phi_{i,m}^T\theta(n)\ge\phi_{i,m}^T(\theta_{i,m}-\theta(n))\big]}\\
\le&\mathbb{I}_{\big[\|y_{i,m+1}-\phi_{i,m}^T\theta(n)\|\le\|\phi_{i,m}^T(\theta_{i,m}-\theta(n))\|\big]}\label{73}
\end{align}
and
\begin{align}\nonumber
&\|\theta_{i,m}-\theta(n)\|\\ \nonumber
\le&\|\theta_{i,m}-\theta_{n_k}\|+\|\theta_{n_k}-\bar{\theta}\|+\|\bar{\theta}-\theta(n)\|\\ \nonumber
\le&\|\theta_{i,m}-\theta_{m}\|+\|\theta_{m}-\theta_{n_k}\|+\|\theta_{n_k}-\bar{\theta}\|+\|\bar{\theta}-\theta(n)\|\\ 
\le&\|\theta_{i,m}-\theta_{m}\|+c_2T_j+\lambda_{n_k}+\lambda_n,\label{74}
\end{align}
where the last inequality follows from Lemma \ref{Lemma4} and the fact that $\|\theta_{n_k}-\bar{\theta}\|<\frac{\lambda_{n_k}}{2}$ and $\|\theta(n)-\bar{\theta}\|<\frac{\lambda_n}{2}$.

Similar to (\ref{59}) we see that there exist positive numbers $c_3$, $c_4$, $c_5$, and $\rho_2\in(0,1)$ such that
\begin{gather}
\|\Theta_{\perp,s+1}\|\le c_3\rho_2^{s+1-n_k}+c_4\sup_{m\ge n_k}a_m+c_5T_j\label{75}
\end{gather}
for sufficiently large $k$ and $\forall s:n_k\le s\le m(n_k,T_j)$.
Since $0<\rho_2<1$, for any fixed $T_j>0$, there exists an integer $m^{\prime}>0$ such that $\rho_2^{m^{\prime}}<T_j$. Since $m(n_k,T_j)-n_k\to\infty$ as $k\to\infty$, we have $n_k+m^{\prime}<m(n_k,T_j)$ for all sufficiently large $k$, and
\begin{align}\nonumber
\|\Theta_{\perp,s+1}\|\le& c_4\sup_{m\ge n_k}a_m+(c_3+c_5)T_j\\
\le& c_4a_{n_k}+(c_3+c_5)T_j\label{76}
\end{align}
for $n_k+m^{\prime}\le s\le m(n_k,T_j)$.

We now consider $\epsilon^{(1)}_{i,k+1}(n)$. From (\ref{72}) it follows that
\begin{align}\nonumber
&\frac{1}{T_j}\Big\|\sum_{m=n_k}^{m(n_k,T_j)}\frac{1}{m}\epsilon_{i,m+1}^{(1)}(n)\Big\|\\
\nonumber
\le&\frac{2}{T_j}\sum_{m=n_k}^{n_k+m^{\prime}}\frac{1}{m}\|\phi_{i,m}\|
+\frac{2}{T_j}\sum_{m=n_k+m^{\prime}}^{m(n_k,T_j)}\frac{1}{m}\|\phi_{i,m}\|\\
&\cdot\mathbb{I}_{\big[\|y_{i,m+1}-\phi_{i,m}^T\theta(n)\|\le\|\phi_{i,m}^T(\theta_{i,m}-\theta(n))\|\big]}.\label{77}
\end{align}


Noticing that the integer $m^{\prime}$ does not depend on $k$ and $\|\phi_{i,k}\|$ is bounded, we conclude that
\begin{gather}
\limsup_{k\to\infty}\sum_{m=n_k}^{n_k+m^{\prime}}\frac{1}{m}||\phi_{i,m}||=0.\label{78}
\end{gather}
We now focus on the second part of (\ref{77}). From (\ref{74}) and (\ref{76}) we have the following chain of equalities and inequalities,
\begin{align}\nonumber
&\frac{2}{T_j}\sum_{m=n_k+m^{\prime}}^{m(n_k,T_j)}\frac{1}{m}\|\phi_{i,m}\| \cdot\mathbb{I}_{\big[\|y_{i,m+1}\!-\!\phi_{i,m}^T\theta(n)\|\le\|\phi_{i,m}^T(\theta_{i,m}\!-\!\theta(n))\|\big]}\\ \nonumber
&\le\frac{2}{T_j}\sum_{m=n_k+m^{\prime}}^{m(n_k,T_j)}\frac{1}{m}\|\phi_{i,m}\|
\cdot\mathbb{I}_{\big[\|y_{i,m+1}\!-\!\phi_{i,m}^T\theta(n)\|\le\|\phi_{i,m}\|\cdot\|\theta_{i,m}\!-\!\theta(n)\|\big]}\\ \nonumber
&\le\frac{2}{T_j}\sum_{m=n_k+m^{\prime}}^{m(n_k,T_j)}\frac{1}{m}\|\phi_{i,m}\|\\ \nonumber
&\cdot\mathbb{I}_{\big[\|y_{i,m+1}-\phi_{i,m}^T\theta(n)\|\le\|\phi_{i,m}\|\cdot \big((c_2+c_3+c_5)T_j+c_4a_{n_k}+\lambda_{n_k}+\lambda_n\big)\big]}
\end{align}
\begin{align}
\nonumber&=\frac{2}{T_j}\sum_{m=n_k+m^{\prime}}^{m(n_k,T_j)}\frac{1}{m}\Big\{\|\phi_{i,m}\|\\ \nonumber
&\cdot\mathbb{I}_{\big[\|y_{i,m+1}-\phi_{i,m}^T\theta(n)\|\le\|\phi_{i,m}\|\cdot \big((c_2+c_3+c_5)T_j+c_4a_{n_k}+\lambda_{n_k}+\lambda_n\big)\big]}\\
\nonumber&-\mathbb{E}\Big(\|\phi_{i,m}\|\\ \nonumber
&\cdot\mathbb{I}_{\big[\|y_{i,m+1}-\phi_{i,m}^T\theta(n)\|\le\|\phi_{i,m}\|\cdot \big((c_2+c_3+c_5)T_j+c_4a_{n_k}+\lambda_{n_k}+\lambda_n\big)\big]}\Big)\Big\}\\
\nonumber&+\frac{2}{T_j}\sum_{m=n_k+m^{\prime}}^{m(n_k,T_j)}\frac{1}{m}\mathbb{E}\Big(\|\phi_{i,m}\|\\
&\cdot\mathbb{I}_{\big[\|y_{i,m+1}-\phi_{i,m}^T\theta(n)\|\le\|\phi_{i,m}\|\cdot \big((c_2+c_3+c_5)T_j+c_4a_{n_k}+\lambda_{n_k}+\lambda_n\big)\big]}\Big).\label{79}
\end{align}

Similar to (\ref{48}), we have that
\begin{align}
\nonumber&\limsup\limits_{k\to\infty}\Big|\frac{2}{T_j}\sum_{m=n_k+m^{\prime}}^{m(n_k,T_j)}\frac{1}{m}\Big\{\|\phi_{i,m}\| \\ \nonumber&\cdot\mathbb{I}_{\big[\|y_{i,m+1}-\phi_{i,m}^T\theta(n)\|\le\|\phi_{i,m}\|\cdot \big((c_2+c_3+c_5)T_j+c_4a_{n_k}+\lambda_{n_k}+\lambda_n\big)\big]}\\
\nonumber&-\mathbb{E}\Big(\|\phi_{i,m}\|\\
\nonumber&\cdot\mathbb{I}_{\big[\|y_{i,m+1}\!-\!\phi_{i,m}^T\theta(n)\|\le\|\phi_{i,m}\|\cdot \big((c_2\!+\!c_3\!+\!c_5)T_j\!+\!c_4a_{n_k}\!+\!\lambda_{n_k}\!+\!\lambda_n\big)\big]}\Big)\Big\}\Big|\\
&=0.\label{80}
\end{align}

Since $\{\phi_{i,k}\}_{k\ge 0}$ is strictly stationary and bounded, for the last term in (\ref{79}) we have
\begin{align}
\nonumber
&\mathbb{E}\Big(\|\phi_{i,m}\|\\
\nonumber&\cdot\mathbb{I}_{\big[\|y_{i,m+1}-\phi_{i,m}^T\theta(n)\|\le\|\phi_{i,m}\|\cdot \big((c_2+c_3+c_5)T_j+c_4a_{n_k}+\lambda_{n_k}+\lambda_n\big)\big]}\Big)\\
\nonumber
&\le C\mathbb{E}\mathbb{I}_{\big[\|y_{i,m+1}-\phi_{i,m}^T\theta(n)\|\le\|\phi_{i,m}\|\cdot \big((c_2+c_3+c_5)T_j+c_4a_{n_k}+\lambda_{n_k}+\lambda_n\big)\big]}\\
\nonumber&=C\int_{\mathbb{R}^l}\int_{\mathbb{R}}\mathbb{I}_{\big[\|s^T(\theta^*-\theta(n))+t\|\le\|s\|\cdot \big((c_2+c_3+c_5)T_j+c_4a_{n_k}+\lambda_{n_k}+\lambda_n\big)\big]}\\
&~~~~\cdot q_i(s) f_{i,d}(t)\mathrm{d}s\mathrm{d}t,\label{81}
\end{align}
which, by the dominated convergence theorem, converges to zero by letting $k\to\infty$, then $j\to\infty$, and finally $n\to\infty$. From (\ref{81}) and noticing $\sum_{m=n_k+m^{\prime}}^{m(n_k,T_j)}\frac{1}{m}<T_j$, we have
\begin{align}
\nonumber&\lim_{j\to\infty}\limsup_{k\to\infty}\frac{1}{T_j}\sum_{m=n_k+m^{\prime}}^{m(n_k,T_j)}\frac{1}{m}
\mathbb{E}\Big(\|\phi_{i,m}\|\\
&\cdot\mathbb{I}_{\big[\|y_{i,m+1}\!-\!\phi_{i,m}^T\theta(n)\|\le\|\phi_{i,m}\|\cdot \big((c_2\!+\!c_3\!+\!c_5)T_j\!+\!c_4a_{n_k}\!+\!\lambda_{n_k}\!+\!\lambda_n\big)\big]}\Big)=0.\label{82}
\end{align}

Combining (\ref{70}), (\ref{78}), (\ref{80}), and (\ref{82}), we obtain that
\begin{align}
&\lim_{j\to\infty}\limsup_{k\to\infty} \frac{1}{T_j}\Big\|\sum_{m=n_k}^{m(n_k,T_j)}\frac{1}{m}\epsilon_{i,m+1}^{(1)}(n)\Big\|=0.\label{83}
\end{align}

For $\epsilon^{(2)}_{i,m+1}(n)$, similar to (\ref{48}) we can prove that
\begin{align}
\nonumber
&\sum_{k=1}^{\infty}\frac{1}{k}\Big[\phi_{i,k}\text{sgn}(y_{i,k+1}-\phi_{i,k}^T\theta(n)) \\ &-\mathbb{E}[\phi_{i,k}\text{sgn}(y_{i,k+1}-\phi_{i,k}^T\theta(n))])\Big]<\infty~~\mathrm{a.s.}~~\forall n\geq1.\label{84}
\end{align}
From this by the definition of $\epsilon^{(2)}_{i,m+1}(n)$, it follows that
\begin{gather}
\lim_{j\to\infty}\limsup_{k\to\infty}\frac{1}{T_j}
\|\sum_{m=n_k}^{m(n_k,T_j)}\frac{1}{m}\epsilon^{(2)}_{i,m+1}\|=0.\label{85}
\end{gather}

For $\epsilon^{(3)}_{i,m+1}(n)$, similar to (\ref{72}), (\ref{73}), and (\ref{82}), we obtain that
\begin{align}\nonumber
&\frac{1}{T_j}\Big\|\sum_{m=n_k}^{m(n_k,T_j)}\frac{1}{m}\epsilon^{(3)}_{i,m+1}(n)\Big\|\\
\nonumber
\le&\frac{2}{T_j}\sum_{m=n_k}^{m(n_k,T_j)}\frac{1}{m}\mathbb{E}\Big\{\|\phi_{i,1}\| \mathbb{I}_{\big[\|y_{i,2}-\phi_{i,1}\bar{\theta}\|\le\|\phi_{i,1}\|\cdot\|\theta(n)-\bar{\theta}\|\big]}\Big\}\\
\le&2\mathbb{E}\Big\{\|\phi_{i,1}\|\mathbb{I}_{\big[\|y_{i,2}-\phi_{i,1}\bar{\theta}\| \le\|\phi_{i,1}\|\cdot\|\theta(n)-\bar{\theta}\|]}\Big\}\to0\label{86}
\end{align}
by letting first $k\to\infty$ and then $n\to\infty$. From (\ref{86}) it follows that
\begin{gather}
\lim_{j\to\infty}\limsup_{k\to\infty}\frac{1}{T_j}
\|\sum_{m=n_k}^{m(n_k,T_j)}\frac{1}{m}\epsilon^{(3)}_{i,m+1}\|=0.\label{87}
\end{gather}

Finally, for $\epsilon^{(4)}_{i,m+1}$, carrying out a treatment similar to that for (\ref{77}), we can prove that
%
%
%
\begin{gather}
\lim_{j\to\infty}\limsup_{k\to\infty}
\frac{1}{T_j}\Big\|\sum_{m=n_k}^{m(n_k,T_j)}\frac{1}{m}\epsilon^{(4)}_{i,m+1}\Big\|=0.\label{88}
\end{gather}

Combining (\ref{83}), (\ref{85}), (\ref{87}), and (\ref{88}) leads to (\ref{66}).\hfill$\square$

The next lemma shows that the truncation number of the distributed identification algorithm is finite, and hence the estimate sequence $\{\theta_{i,k}\}_{k\ge 0}$ is bounded for any agent $i\in \mathcal{V}$.

\begin{lemma} \label{Lemma6}
If C1)--C4) hold, then
\begin{gather}
\lim_{k\to\infty}\sigma_k=\sigma<\infty~~\mathrm{a.s.}\label{89}
\end{gather}
\end{lemma}

\emph{\textbf{Proof}}: From Lemma \ref{Lemma3} we know that the estimate sequence $\{\Theta_{k}\}$ generated by (\ref{6})--(\ref{5}) contains a bounded subsequence $\{\Theta_{n_k}\}$ with $\sigma_{i,n_k}=\sigma_{n_k}~\forall i\in\mathcal{V}$. For this bounded subsequence $\{\Theta_{n_k}\}_{k\ge1}$, there exists a constant $c_0$ such that $\|\Theta_{n_k}\|\le c_0$. Thus, $\{\theta_{n_k}\}$ is also located in the bounded set $\{\theta\in\mathbb{R}^l:\|\theta\|\le c_0\}$.

Set $v(\theta)\triangleq\mathbb{E}\|y_2-\phi_1^T\theta\|_1$. Since $v(\theta)$ is convex, there exists a positive constant $c_1>c_0$ such that $\max_{||\theta||<c_0}v(\theta)<\inf_{||\theta||=c_1}v(\theta)$. Since in Lemma \ref{Lemma1} it is shown that $J=\{\theta^*\}$, there exists a nonempty interval $[\delta_1,\delta_2]\in(\max_{||\theta||<c_0}v(\theta),\inf_{||\theta||=c_1}v(\theta))$ such that $d([\delta_1,\delta_2],v(J))>0$.

We now prove (\ref{89}).

Assume the converse that $\lim_{k\to\infty}\sigma_{k}=\infty$. Carrying out a treatment similar to the proof of Lemma 5.4 in \cite{lei2019distributed}, we can prove that $\{\theta_{n_k}\}$ starting from a point in the set $\{\theta\in\mathbb{R}^l:\|\theta\|\le c_0\}$ crosses the boundary $\{\theta\in\mathbb{R}^l:\|\theta\|=c_1\}$ infinitely many times. Therefore, for the nonempty interval $[\delta_1,\delta_2]$, there are infinitely many crossings $\{v(\theta_{n_k}),\dots,v(\theta_{m_k})\}$. Here by ``crossing $[\delta_1,\delta_2]$ by $\{v(\theta_{n_k}),\dots,v(\theta_{m_k})\}$" we mean that $v(\theta_{n_k})\le\delta_1,v(\theta_{m_k})\ge\delta_2$, and $\delta_1<v(\theta_s)<\delta_2~\forall s:n_k<s<m_k$.

Set
$$
O_{i,k+1}=f_i(\theta_{i,k})+\epsilon_{i,k+1}.
$$

So, the algorithm (\ref{6})--(\ref{5}) is a DSAAWET given in Appendix. Since $a_k=\frac{1}{k}$, A1) in Theorem \ref{TheoremA1} in Appendix  is satisfied. Noticing $f_i(\theta)=\mathbb{E}[\phi_{i,k}\text{sgn}(y_{i,k+1}-\phi_{i,k}^T\theta)]~\forall i\in\mathcal{V}$, by (\ref{14}) and C3), we see that $f_i(\theta)$ is continuous. Hence A3) required by Theorem \ref{TheoremA1} in Appendix holds true. Since $f(\theta)=\sum_{i=1}^N\mathbb{E}[\phi_{i,k}\text{sgn}(y_{i,k+1}-\phi_{i,k}^T\theta)]$ and $J=\{\theta^*\}$, by setting $v(\theta)=\mathbb{E}||y_{k+1}-\phi_k^T\theta||_1$ it is seen that A2) is satisfied. Further, in Lemma \ref{Lemma5}, we have proved that the noise sequence satisfies (\ref{66}) along the indices $\{n_k\}$ of any convergent subsequence $\{\Theta_{n_k}\}$ with $\sigma_{i,n_k}=\sigma_{n_k}~\forall i\in\mathcal{V}$. Then similar to the proof of Lemma 5.3 in \cite{lei2019distributed}, we can show that any nonempty interval $[\delta_1,\delta_2]$ with $d([\delta_1,\delta_2],v(J))>0$ cannot be crossed by infinitely many sequences $\{v(\theta_{n_k}),\dots,v(\theta_{m_k})\}$. This yields a contradiction.

Thus, the number of truncations must be finite and hence (\ref{89}) holds.\hfill$\square$

The following lemma shows that consensus of the distributed identification algorithms can be achieved.

\begin{lemma} \label{Lemma7} (Consensus of Estimates)
If C1)--C4) hold, then
\begin{gather}
\|\Theta_{\perp,k}\|\mathop{\longrightarrow}0~~\mathrm{as}~~k\to\infty~~\mathrm{a.s.}\label{90}
\end{gather}
\end{lemma}

\emph{\textbf{Proof}}: From Lemma \ref{Lemma6} we know that, for (\ref{6})--(\ref{5}) the number of truncations is finite. Then by iii) in Lemma \ref{Lemma2} it follows that there exists a positive integer $\sigma$ such that $\hat{\sigma}_{i,k}=\sigma_{i,k}=\sigma$ for any $ k\ge k_0=BD+\tau_{\sigma}$ and any $i\in \mathcal{V}$. So for any $k\ge k_0$, the algorithm (\ref{6})--(\ref{5}) can be rewritten as:
\begin{gather}
\Theta_{k+1}=(W(k)\otimes\textbf{I}_l)\Theta_k+\frac{1}{k}(F(\Theta_k)+\epsilon_{k+1}).\label{91}
\end{gather}

Pre-multiplying both sides of (\ref{91}) with $D_{\perp}\triangleq (\textbf{I}_N-\frac{\textbf{1}\textbf{1}^T}{N})\otimes\textbf{I}_l$, we obtain that
\begin{gather}
\Theta_{\perp,k+1}=D_{\perp}(W(k)\otimes\textbf{I}_l)\Theta_k +\frac{1}{k}D_{\perp}(F(\Theta_k)+\epsilon_{k+1}).\label{92}
\end{gather}

Set
\begin{align}
\nonumber
&\Psi(k,s)\triangleq[D_{\perp}(W(k)\otimes\textbf{I}_l)]\cdots
[D_{\perp}(W(s)\otimes\textbf{I}_l)]~~\forall k\ge s,\\ \nonumber~~&\Psi(k-1,k)\triangleq\textbf{I}_{Nl}.
\end{align}

Since $\{W(k)\}_{ k\ge 1}$ are doubly stochastic, it directly follows that
\begin{align}
\nonumber
&\Psi(k,s)=(\Phi(k,s)-\frac{1}{N}\textbf{1}\textbf{1}^T)\otimes\textbf{I}_l,\\
\nonumber
&\Psi(k,s)D_{\perp}=(\Phi(k,s)-\frac{1}{N}\textbf{1}\textbf{1}^T)\otimes\textbf{I}_l~~\forall k\ge s.
\end{align}
By this and (\ref{92}), we have
\begin{align}
\nonumber
&\Theta_{\perp,k+1}\\
\nonumber=&\Psi(k,k_0)\Theta_{k_0}+\sum_{m=k_0}^k\frac{1}{m}\Psi(k-1,m)D_{\perp}(F(\Theta_k)+\epsilon_{k+1})\\
\nonumber
=&[(\Phi(k,k_0)-\frac{1}{N}\textbf{1}\textbf{1}^T)\otimes\textbf{I}_l]\Theta_{k_0}\\ \nonumber
&+\sum_{m=k_0}^k\frac{1}{m}[(\Phi(k-1,m)-\frac{1}{N}\textbf{1}\textbf{1}^T)\otimes\textbf{I}_l]F(\Theta_m)\\
&+\sum_{m=k_0}^k\frac{1}{m}[(\Phi(k-1,m)-\frac{1}{N}\textbf{1}\textbf{1}^T)\otimes\textbf{I}_l]\epsilon_{m+1}.\label{93}
\end{align}

By the continuity of $F(\theta)$ and the boundedness of $\{\Theta_k\}$ established in Lemma \ref{Lemma6}, and by noticing (\ref{8}), it follows that there exist positive constants $c_1^{\prime}$, $c_2^{\prime}$, $c_3^{\prime}$ and $0<\rho_2<1$ such that for all $ k\ge k_0$
\begin{align}\nonumber
\|\Theta_{\perp,k+1}\|\le& c_1^{\prime}\rho_2^{k-1+k_0}+c_2^{\prime}\sum_{m=k_0}^k\frac{1}{m}\rho_2^{k-m}\\
&+c_3^{\prime}\sum_{m=k_0}^k\frac{1}{m}\rho_2^{k-m}\|\epsilon_{m+1}\|.\label{94}
\end{align}

Noticing that $0<\rho_2<1$ and $\epsilon_{k}$ is bounded, we have that
\begin{align}
&\sum_{m=k_0}^k\frac{1}{m}\rho_2^{k-m}\mathop{\longrightarrow}\limits_{k\to\infty}0,\label{95}
\end{align}
and
\begin{align}
&\sum_{m=k_0}^k\frac{1}{m}\rho_2^{k-m}\|\epsilon_{m+1}\|=O\left(\sum_{m=k_0}^k\frac{1}{m}\rho_2^{k-m}\right) \mathop{\longrightarrow}\limits_{k\to\infty}0.\label{96}
\end{align}

From (\ref{94})--(\ref{96}), we conclude that
$
\Theta_{\perp,k}\mathop{\longrightarrow}\limits_{k\to\infty}0.
$
This finishes the proof.\hfill$\square$

We now prove the strong consistency of the estimates generated by (\ref{6})--(\ref{5}).

\emph{\textbf{Proof of Theorem \ref{Theorem1}}}: For (\ref{97}) we only need to show that A1)--A5) required by Theorem \ref{TheoremA1} in Appendix hold true.

In the proof of Lemma \ref{Lemma6} we have verified A1), A2), and A3). Note that C4) coincides with A5) given in Appendix. Thus, it remains to verify A4).

By the boundedness of $\{\epsilon_{i,k+1}\}$, it follows that
\begin{align}
\lim_{k\to\infty}\frac{1}{k}\epsilon_{i,k+1}=0.\label{98}
\end{align}
So A4) a) is satisfied. We now verify A4) b).

For agent $i$, denote by $\{\theta_{i,n_k}\}$ any convergent subsequence of $\{\theta_{i,k}\}$. By Lemma \ref{Lemma6}, we have shown that along indices $\{n_k\}$ the estimate sequence $\{\Theta_{n_k}\}$ is bounded, and for sufficiently large $k$ we have $\sigma_{j,n_k}=\sigma_{n_k}=\sigma,~j\in \mathcal{V}$. So, (\ref{31}) and (\ref{32}) in Lemma \ref{Lemma4} can be applied for the indices $\{n_k\}$ considered here.

Similar to Lemma \ref{Lemma5}, denote by $\bar{\theta}$ the limit of $\{\theta_{i,n_k}\}$, and $\mathcal{S}$ a countable set dense in $\mathbb{R}^l$. Let $\{\theta(n)\}_{n\ge 1}\subset\mathcal{S}$ be a sequence tending to $\bar{\theta}$ such that $\|\theta_{i,n_k}-\bar{\theta}\|\leq\|\theta(n_k)-\bar{\theta}\|$. Let $\{\lambda_n\}\subset\mathcal{S}$ be a sequence of positive numbers with $\lambda_n\to 0$ as $n\to\infty$ such that $||\theta(n)-\bar{\theta}||<\frac{\lambda_n}{2}$. Let $\{T_j\}_{j\ge1}\subset\mathcal{S}$ be a sequence of positive numbers tending to zero with $T_j>T_{j+1}$. We rewrite $\epsilon_{i,m+1}$ as
\begin{gather}
\epsilon_{i,m+1}\!=\!\epsilon_{i,m+1}^{(1)}(n)\!+\!\epsilon_{i,m+1}^{(2)}(n)\! +\!\epsilon_{i,m+1}^{(3)}(n)\!+\!\epsilon_{i,m+1}^{(4)},\label{99}
\end{gather}
where
\begin{align}\nonumber
\epsilon_{i,m+1}^{(1)}(n)=&\phi_{i,m}\text{sgn}(y_{i,m+1}-\phi_{i,m}^T\theta_{i,m})\\
&-\phi_{i,m}\text{sgn}(y_{i,m+1}-\phi_{i,m}^T\theta(n)),\label{100}\\\nonumber
\epsilon_{i,m+1}^{(2)}(n)=&\phi_{i,m}\text{sgn}(y_{i,m+1}-\phi_{i,m}^T\theta(n))\\
&-\mathbb{E}[\phi_{i,m}\text{sgn}(y_{i,m+1}-\phi_{i,m}^T\theta(n))],\label{101}\\ \nonumber
\epsilon_{i,m+1}^{(3)}(n)=&\mathbb{E}[\phi_{i,m}\text{sgn}(y_{i,m+1}-\phi_{i,m}^T\theta(n))]\\
&-\mathbb{E}[\phi_{i,m}\text{sgn}(y_{i,m+1}-\phi_{i,m}^T\bar{\theta})],\label{102}
\end{align}
and
\begin{align}
\nonumber
\epsilon_{i,m+1}^{(4)}=&\mathbb{E}[\phi_{i,m}\text{sgn}(y_{i,m+1}-\phi_{i,m}^T\bar{\theta})]\\
&-\mathbb{E}[\phi_{i,m}\text{sgn}(y_{i,m+1}-\phi_{i,m}^T\theta_{i,m})].\label{103}
\end{align}

The following proof is similar to that of Lemma \ref{Lemma5}. Here we only sketch the proof for $\epsilon^{(1)}_{i,m+1}(n)$.

For $m=n_k,\cdots,m(n_k,T_j)$, we have the following inequality,
\begin{align}\nonumber
&\|\theta_{i,m}-\theta(n)\|\\ \nonumber
\le&\|\theta_{i,m}-\theta_m\|+\|\theta_m-\theta_{n_k}\|+\|\theta_{n_k}-\theta_{i,n_k}\|\\
&+\|\theta_{i,n_k}-\bar{\theta}\|+\|\bar{\theta}-\theta(n)\|.\label{104}
\end{align}

By Lemma \ref{Lemma7} we know that $\|\theta_{i,k}-\theta_{k}\|\mathop{\longrightarrow}\limits_{k\to\infty}0$. Let $\{\gamma_k\}\subset\mathcal{S}$ be a sequence of positive numbers tending to zero as $k\to\infty$ such that $\|\theta_{i,k}-\theta_{k}\|<\frac{\gamma_k}{2}$. From (\ref{104}) and (\ref{32}), we obtain that
\begin{align}
\|\theta_{i,m}-\theta(n)\|
\le&\gamma_m+c_2T_j+\gamma_{n_k}+\lambda_{n_k}+\lambda_n.\label{105}
\end{align}


By using the same analysis as that for Lemma \ref{Lemma5} we can prove that
\begin{gather}\nonumber
\lim_{j\to\infty}\limsup_{k\to\infty}\frac{1}{T_j}\|\sum_{m=n_k}^{m(n_k,T_j)}\frac{1}{m}\epsilon^{(1)}_{i,m+1}(n)\|=0,
\end{gather}
and further,
\begin{gather}\nonumber
\lim_{j\to\infty}\limsup_{k\to\infty}\frac{1}{T_j}\|\sum_{m=n_k}^{m(n_k,T_j)}\frac{1}{m}\epsilon_{i,m+1}\|=0~~\forall i\in \mathcal{V},
\end{gather}
which implies A4) b).

Then by Theorem \ref{TheoremA1} in Appendix it follows that the estimates generated by (\ref{6})--(\ref{5}) converge to $\theta^*$.\hfill$\square$

\section{Numerical simulation}

Consider a network $\mathcal{G}=(\mathcal{V},\mathcal{E})$ with $\mathcal{V}=\{1,\cdots,N\},~N=100,$ and $\mathcal{E}=G(N,p_N)$ being the Poisson random graph\footnote{For the details of Poisson random graph, we refer to \cite{jackson2010social}.} with designing parameter $0\leq p_N\leq 1$. We choose $p_N=6/N$. Denote by $N_i$ the neighbor set of agent $i$ and by $n_i$ the cardinality of $N_i$. Set $W(k)=[w_{ij}]_{i,j=1}^N~\forall k\ge1$ with $w_{ij}=\frac{1}{n_i}$ if agent $j$ is in the set $N_i$. The dynamics of each agent $i\in \mathcal{V}$ is given by
\begin{align*}
y_{i,k+1}=\phi_{i,k}^T\theta^*+d_{i,k+1},~~z_{i,k+1}=\mathbb{I}_{[y_{i,k+1}<c_{i,k}]},
\end{align*}
where $y_{i,k}\in \mathbb{R}^1,~\theta^*\in \mathbb{R}^l,~l=8$ and the $j$-th entry of $\theta^*$ is $(1+0.1j)\sqrt{j}$.

Let $\{\eta_{i,k}\}_{k\geq1},~i=1,\cdots,N$ be sequences of i.i.d. random variables uniformly distributed over $[-1,1]$ with $\{\eta_{i,k}\}_{k\geq1}$ and $\{\eta_{j,k}\}_{k\geq1}$ being mutually independent if $i\neq j$. At time $k$, for the regressor $\phi_{i,k}\in\mathbb{R}^{l}$, if $i\mod l\neq0$, then its $i\mod l$-th entry is set to be $\eta_{i,k}$ and other entries are set to be $0$, while if $i\mod l=0$, then its $l$-th entry is set to be $\eta_{i,k}$ and other entries are set to be $0$. Assume that $\{d_{i,k}\}_{k\ge1}$, $i\in\mathcal{V}$ are sequences of i.i.d. random variables with Gaussian distribution $\mathcal{N}(0,0.09)$ and $\{\eta_{i,k}\}_{k\ge1}$ and $\{d_{j,k+1}\}_{k\ge1}$ are mutually independent for $i\neq j$.

Denote by $\{\theta_{i,k}\}_{k\ge1},i\in \mathcal{V}$ the estimates given by (\ref{6})--(\ref{5}) and by $\theta_k=\frac{1}{N}\sum_{i=1}^N\theta_{i,k}$ the average of $\theta_{i,k},~i\in \mathcal{V}$. In Figure \ref{Figure1}, the dashed lines denote the true values of parameters and the solid lines the estimates for entries $\{\theta_k^j,~j=1,\cdots,l\}_{k\geq1}$ of $\{\theta_k\}_{k\geq1}$. From the figure we find that the simulation results are consistent with the theoretical analysis.
\begin{figure}[htp]
 \centering
 \includegraphics[width=9cm]{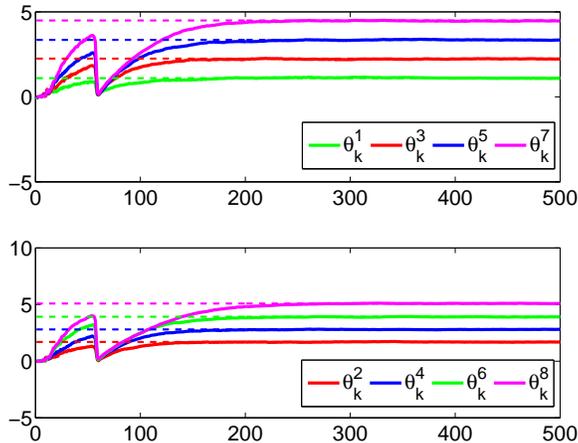}
 \caption{Estimation sequences of $\theta_k^j,j=1,\cdots,l$.}
 \label{Figure1}
\end{figure}

\section{Concluding Remarks}

The distributed parameter estimation of linear stochastic system over time-varying networks with binary sensors is considered in the paper. Each agent in the network can only access the input as well as the binary-valued output of the local system, but aims at estimating the global unknown parameters. A DSAAWET-based identification algorithm is proposed and the consensus and convergence of the estimates are established.

For future research, it is of interest to relax the technical assumptions adopted in this paper, in particular, the boundedness assumption on the regressors. It is also of interest to consider the distributed identification of nonlinear stochastic systems.

\section*{Appendix}



For a time-varying network $(\mathcal{V},\mathcal{E}(k)),~k\geq1$ with $\mathcal{V}=\{1,\cdots,N\}$, consider the distributed root-searching of
$
f(x)=\sum_{i=1}^Nf_i(x),
$
on the basis of local observation $f_i(\cdot):\mathbb{R}^l\to\mathbb{R}^l$ of agent $i\in \mathcal{V}$ and the information obtained from its adjacent neighbours.

Denote by $J\triangleq\{x\in\mathbb{R}^l:f(x)=0\}$ the root set of $f(\cdot)$ and by $x_{i,k}\in\mathbb{R}^l$ the estimate for the root of $f(\cdot)$ generated by agent $i$ at time $k$. The local observation of agent $i$ is given by
\begin{gather}
O_{i,k+1}=f_{i}(x_{i,k})+\epsilon_{i,k+1},\label{106}
\end{gather}
where $\epsilon_{i,k+1}$ is the observation noise. With $\{M_k\}$ being a sequence of positive numbers increasingly diverging to infinity and $x^*\in\mathbb{R}^l$ being a given point known to all agents, the estimates $\{x_{i,k}\}_{k\ge1}$ at agent $i$ are generated as follows:
\begin{align}
\sigma_{i,0}=&0,~\hat{\sigma}_{i,k}=\max_{j\in N_i(k)}\sigma_{j,k},\label{107}\\
\nonumber
x_{i,k+1}^{\prime}=&\{\sum_{j\in N_i(k)}w_{ij}(k)(x_{j,k}\mathbb{I}_{[\sigma_{j,k}=\hat{\sigma}_{i,k}]} +x^*\mathbb{I}_{[\sigma_{j,k}<\hat{\sigma}_{i,k}]})\\
&+a_kO_{i,k+1}\}\cdot\mathbb{I}_{[\sigma_{i,k}=\hat{\sigma}_{i,k}]} +x^*\mathbb{I}_{[\sigma_{i,k}<\hat{\sigma}_{i,k}]},\label{108}\\
x_{i,k+1}=&x_{i,k+1}^{\prime}\mathbb{I}_{[||x_{i,k+1}^{\prime}||\le M_{\hat{\sigma}_{i,k}}]}+x^*\mathbb{I}_{[||x_{i,k+1}^{\prime}||> M_{\hat{\sigma}_{i,k}]}},\label{109}\\
\sigma_{i,k+1}=&\hat{\sigma}_{i,k}+\mathbb{I}_{[||x_{i,k+1}^{\prime}||>M_{\hat{\sigma}_{i,k}}]},\label{110}
\end{align}
where $a_k$ is the step size.

The following conditions are used:
\begin{itemize}

\item[A1)] $a_k>0,a_k\to 0,\sum_{k=1}^{\infty}a_k=\infty$.

\item[A2)] There exists a continuously differentiable function $v(\cdot):\mathbb{R}^l\to\mathbb{R}$ such that
$
\sup_{r_1\le \mathrm{d}(x,J)\le r_2}f^T(x)v_x(x)<0
$
for any $0<r_1<r_2<\infty$, where $v_x(\cdot)$ denotes the gradient of $v(\cdot)$ and $\mathrm{d}(x,J)=\inf_{y}\{\|x-y\|:y\in J\}$ and $v(J)\triangleq\{v(x):x\in J\}$ is nowhere dense. Further, $x^*$ adopted in (\ref{108}) and (\ref{109}) satisfies that $||x^*||<c_0$ and $v(x^*)<\inf_{||x||=c_0}v(x)$ for some positive constant $c_0$.

\item[A3)] The local functions $f_i(\cdot)~~\forall i\in\mathcal{V}$ are continuous.

\item[A4)] For any $i\in\mathcal{V}$, the noise sequence $\{\epsilon_{i,k+1}\}_{k\ge 0}$ satisfies\\
a) $\lim_{k\to\infty}a_k\epsilon_{i,k}=0$;\\
b) $\lim_{T\to 0}\limsup_{k\to\infty}\frac{1}{T}||\sum_{m=n_k}^{m(n_k,t_k)}a_m\epsilon_{i,m}||=0$ for any $t_k\in[0,T]$, where $m(k,T)\triangleq\max\{m:\sum_{i=k}^ma_i\le T\}$ and
$\{n_k\}$ denotes the indices of any convergent subsequence of $\{x_{i,k}\}$.

\item[A5)] For the time-varying network $(\mathcal{V},\mathcal{E}(k))$, the following conditions are assumed.\\
a) The adjacent matrices $W(k)$ are doubly stochastic for each $k\geq 0$;\\
b) There exists a constant $0<\kappa<1$ such that
$
w_{ij}(k)\ge\kappa,
$
whenever $j\in N_i(k)$ for all $i\in\mathcal{V}$ and $k\geq 0$;\\
c) The digraph $\mathcal{G}_{\infty}=\{\mathcal{V},\mathcal{E}_{\infty}\}$ is strongly connected with
$
\mathcal{E}_{\infty}\triangleq\{(j,i):(j,i)\in\mathcal{E}(k)$ for infinitely many $k\},$\\
d) There exists a positive integer $B$ such that
$
(j,i)\in\mathcal{E}(k)\cup\mathcal{E}(k+1)\cup\cdots\cup\mathcal{E}(k+B-1)
$
for any $(j,i)\in\mathcal{E}_{\infty}$ and any $k\ge 1$.
\end{itemize}

\begin{theorem} (\cite[Theorem 3.3]{lei2019distributed}) \label{TheoremA1}\\
Let $\{x_{i,k}\},i\in \mathcal{V}$ be generated by (\ref{107})--(\ref{110}) with any initial value $x_{i,0}$. Assume A1)--A3) and A5) hold. Then
$
X_{\perp,k}\mathop{\longrightarrow}0~\text{and}~d(x_k,J)\mathop{\longrightarrow}0
$ as $k\to\infty$
on the sample path $\omega$ for which A4) holds for all agents, where $x_k=\frac{1}{N}\sum_{i=1}^Nx_{i,k}$, $X_k=\mathrm{col}\{x_{1,k},\dots,x_{N,k}\}$, and $X_{\perp,k}=X_k-(\textbf{1}\otimes\textbf{I}_l)x_k$.
\end{theorem}


\begin{theorem}\label{TheoremA2}(\cite{ZhaoZhengBai})
Assume $\{\phi_k\}_{k\ge 0}$ with $\phi_k\in\mathbb{R}^l$ is an $\alpha$-mixing with mixing coefficients denoted by $\{\alpha(k)\}_{k\ge 0}$. Let $\{H_k(\cdot)\}_{k\ge0}$ be a sequence of functions $H_k(\cdot):\mathbb{R}^l\to\mathbb{R}$ and $\mathbb{E}H_k(\phi_k)=0$. If there exist constants $\epsilon>0$ and $\gamma>0$ such that
$
\sum_{k=1}^{\infty}(\mathbb{E}|H_k(\phi_k)|^{2+\epsilon})^{\frac{2}{2+\epsilon}}<\infty
$
and
$
\sum_{k=1}^{\infty}\log k(\log\log k)^{1+\gamma}(\alpha(k))^{\frac{2}{2+\epsilon}}<\infty,
$
then
$
\sum_{k=1}^{\infty}H_k(\phi_k)<\infty~~\mathrm{a.s.}
$
\end{theorem}

\end{document}